\documentclass[a4paper,fleqn,usenatbib]{mnras}

\usepackage{txfonts}

\usepackage[T1]{fontenc}
\usepackage{ae,aecompl}

\usepackage{graphicx}	
\usepackage{amstext}
\usepackage{amssymb}	
\usepackage{geometry,lscape,afterpage}
\usepackage{longtable}

\newcommand{\lya}{Ly$\alpha$}

\newcommand{\nqso}{28}
\newcommand{\nlls}{52}

\newcommand{\qname}{J124957.23$-$015928.8}

\title[MAGG I.]{MUSE Analysis of Gas around Galaxies (MAGG) - I: Survey design and the environment of a near pristine gas cloud at $z\approx 3.5$.}

\author[Lofthouse et al.]{Emma K. Lofthouse$^{1,2}$\thanks{E-mail: emma.k.lofthouse@durham.ac.uk}, 
  Michele Fumagalli$^{1,2,3}$\thanks{E-mail: michele.fumagalli@durham.ac.uk}, 
  Matteo Fossati$^{1,2}$, John M. O'Meara$^{4}$, \and Michael T. Murphy$^{5}$, Lise Christensen$^{6}$, J. Xavier Prochaska$^{7}$,  Sebastiano Cantalupo$^{8}$, \and Richard M. Bielby$^{2}$, Ryan J. Cooke$^{2}$, Elisabeta Lusso$^{9}$, Simon L. Morris$^{2, 10}$ \\
  $^{1}$Institute for Computational Cosmology, Durham University, South Road, Durham, DH1 3LE, UK \\
  $^{2}$Centre for Extragalactic Astronomy, Durham University, South Road, Durham, DH1 3LE, UK \\
  $^{3}$Dipartimento di Fisica G. Occhialini, Universit\`a degli Studi di Milano Bicocca, Piazza della Scienza 3, 20126 Milano, Italy \\
  $^{4}$W. M. Keck Observatory 65-1120 Mamalahoa Hwy. Kamuela, HI 96743 \\
  $^{5}$Centre for Astrophysics and Supercomputing, Swinburne University of Technology, Hawthorn, Victoria 3122, Australia\\
  $^{6}$DARK, Niels Bohr Institute, University of Copenhagen, Lyngbyvej 2, DK-2100, Copenhagen, Denmark\\
  $^{7}$Department of Astronomy and Astrophysics, University of California, Santa Cruz, CA 95064, USA \\
  $^{8}$ Department of Physics, ETH Zurich, Wolfgang-Pauli-Strasse 27, 8093 Zurich, Switzerland\\
  $^{9}$ Dipartimento di Fisica e Astronomia, Universit\`a degli Studi di Firenze, Via Giovanni Sansone, 1 50019 Sesto Fiorentino, Italy \\
  $^{10}$ Centre for Advanced Instrumentation, Durham University, South Road, Durham, DH1 3LE, UK }

\pubyear{\the\year}

\begin{document}
\label{firstpage}
\pagerange{\pageref{firstpage}--\pageref{lastpage}}
\maketitle
\begin{abstract}
  We present the design, methods, and first results of the MUSE Analysis of Gas around Galaxies (MAGG) survey,   a large programme on the Multi Unit Spectroscopic Explorer (MUSE) instrument at the Very Large Telescope (VLT) which targets 28 $z\textgreater 3.2$ quasars to investigate the connection between optically-thick gas and galaxies at $z\sim 3-4$. MAGG maps the environment of \nlls\ strong absorption line systems at $z \gtrsim 3$, providing the first statistical sample of galaxies associated with gas-rich structures in the early Universe.
  In this paper, we study the galaxy population around a very metal poor gas cloud at $z\approx 3.53$ towards the quasar \qname. We detect three \lya\ emitters within $\lesssim 200~\rm km~s^{-1}$ of the cloud redshift, at projected separations $\lesssim 185~\rm ~kpc$ (physical). The presence of star-forming galaxies near a very metal-poor cloud indicates that metal enrichment is still spatially inhomogeneous at this redshift. Based on its very low metallicity and the presence of nearby galaxies, we propose that the most likely scenario for this LLS is that it lies within a filament which may be accreting onto a nearby galaxy. Taken together with the small number of other LLSs studied with MUSE, the observations to date show a range of different environments near strong absorption systems. The full MAGG survey will significantly expand this sample and enable a statistical analysis of the link between gas and galaxies to pin down the origin of these diverse environments at $z\approx 3-4$.
\end{abstract}




\section{Introduction}

The evolution of galaxies throughout cosmic time is tightly linked 
to the processes that regulate the supply of gas available for the formation of stars. In a cold dark matter (CDM) Universe, galaxies form within matter overdensities that detach from the
Hubble flow and collapse to form halos \citep[e.g.][]{gunn1972}.
Galaxies grow inside these dark matter halos by acquiring gas either through accretion via cooling of a hot gas halo, or directly via cold gas that streams inward along the cosmic web filaments \citep[e.g][]{white1978,blumenthal1984,keres2005,dekel2006}. As this gas is then converted into stars inside the interstellar medium (ISM), the injection of energy and momentum from processes related to stellar evolution, supernovae explosions, and active galactic nuclei regulates what fraction of the cosmologically-accreted baryons is retained inside galaxies or ejected back into the intergalactic medium (IGM), where it contributes to the observed metal enrichment \citep[e.g][]{dekel1986,schaye2003,springel2005,lilly2013}.      

Within this picture, an important factor regulating the the build-up of galaxies as a function of time is the balance between inflows and outflows. Star formation becomes a second-order variable that, on shorter time scales ($\lesssim 1-2~\rm Gyr$) than the Hubble time, converts the gas supply inside galaxies into stars \citep[e.g.][]{bouche2010,dave2012}. Hence, a full appreciation of how inflows and outflows interact and coexist at the boundary between the ISM and the IGM, within the circumgalactic medium \citep[CGM;][]{steidel2010,tumlinson2017}, becomes a key stage for a complete theory of galaxy evolution.
At the same time, environmental processes triggered by the interactions between galaxies themselves and between galaxies and the more diffuse gas locked in halos or within the cosmic web cannot be neglected \citep[e.g.][]{boselli2006}, as they act as an additional variable that regulates the gas supply.

Several dedicated surveys have been undertaken in recent years to advance our view of inflows and outflows in proximity to galaxies at various redshifts. Significant progress has been made especially at $z \lesssim 1$, due to the availability of large spectroscopic surveys (e.g. the Sloan Digital Sky Survey, \citealt{york2001}, or the Galaxy And Mass Assembly survey, \citealt{driver2011}) that, supplemented by follow-up spectroscopy of quasars in the optical and UV, allow for detailed studies of the CGM in absorption as a function of galaxies properties in emission (including mass, star formation rates, and luminosity), and their environment \citep[e.g.][]{tumlinson2013,stocke2013,bordoloi2014,finn2016,kauffmann2017,heckman2017}.
These studies reveal the ubiquitous presence of a multiphase, enriched, and kinematically-complex CGM surrounding every galaxy, containing a significant baryonic mass that is comparable to, or even in excess of, the mass of baryons locked in stars. 

Likewise, there have been significant efforts in understanding the connection between the CGM probed by quasar spectroscopy and galaxies detected in emission at $z \gtrsim 1$ via dedicated observing campaigns made possible by multi-object spectrographs  \citep[e.g.][]{steidel2010,rubin2010,crighton2011,rudie2012,turner2014,tummuangpak2014, bielby2017b}. 
Similarly to what is found at lower redshift, these experiments reveal the presence of a metal-enriched and multiphase CGM near galaxies, with kinematics consistent with the presence of inflows and outflows inside and near halos \citep[e.g.][]{steidel2010,turner2017}. Despite significant advancements, however, our view of the CGM at large cosmic distances has been mostly limited to star-forming galaxies at the bright end of the UV luminosity function. Moreover, it has proven rather
difficult to obtain high-density spectroscopy at close angular separations from the quasar sightlines with multi-object spectrographs  ($\lesssim 25~\rm arcsec$, corresponding to $\lesssim 200~\rm physical~kpc$ at $z\approx 2-3$). Hence, most of the statistical power of current surveys at $z\gtrsim 3$ is on scales of $\approx 0.1-1~\rm Mpc$ around galaxies, with only a handful of systems available for the study of the inner CGM. Finally, the need for pre-selection of targets for spectroscopic follow-up has hampered
a detailed characterisation of the environment near these systems, and particularly of Ly$\alpha$-bright but UV-faint galaxies  \citep[e.g.][]{crighton2015}.

The Multi Unit Spectroscopic Explorer \citep[MUSE;][]{bacon2010} at the ESO Very Large Telescopes (VLT) represents a significant breakthrough for these types of studies, as its $1\times 1~\rm arcmin^2$ field of view (FOV) enables deep spectroscopic surveys of regions of $\approx 500 \times 500~\rm kpc^2$ at $z\approx 3$
to obtain  highly-complete (to a given flux limit) searches of galaxies near quasar sightlines, with the exception of
very dust obscured systems. Thus, as demonstrated by previous studies \citep[e.g.][]{bouche2016,fumagalli2016,fumagalli2017,zabl2019}, MUSE is a very efficient instrument to improve our view of the inner CGM ($\lesssim 200~\rm kpc$) of high-redshift galaxies, and to investigate their  environment  \citep[e.g.][]{bielby2017,peroux2017}. Furthermore, the ability to reconstruct spatially-resolved emission line maps offers the exciting prospect of investigating the denser parts of the CGM in emission \citep[e.g.][]{borisova2016,wisotzki2016,leclercq2017,cai2018,ginolfi2018,arrigoni2018,arrigoni2019}.

Leveraging this technological advancement, we designed the 
MUSE Analysis of Gas around Galaxies (MAGG), a survey which builds on a MUSE Large Programme (ID 197.A$-$0384; PI Fumagalli) to explore the co-evolution of gas and
galaxies in \nqso\ quasar fields at redshift $z \approx 3.2 - 4.5$, for which  high-resolution spectroscopy is available. Our survey is primarily intended to complement previous studies of the CGM of galaxies in the range $z\approx 2.0-4.0$ by focusing on low-mass
galaxies detected via Ly$\alpha$ emission or absorption features. Further, we focus on quasar sightlines that host strong absorption line systems
with hydrogen column densities $N_{\rm HI}\gtrsim 10^{17}~\rm cm^{-2}$, which act as signposts of a dense and (partially) neutral phase inside the CGM \citep[e.g.][]{faucher2011,fumagalli2011,vandevoort2012,fumagalli2013}. The versatile nature of MUSE further allows the study of quasars and their environment, and of galaxies associated with lower-redshift absorbers.

In this first paper of a series, we present the survey strategy and sample selection (Sect.~\ref{sec:selection}), and discuss in
detail the processing of MUSE data and high-resolution spectroscopy (Sect.~\ref{sec:reduction}), including the methodology
adopted to derive catalogues of galaxies and absorbers (Sect.~\ref{sec:als}-\ref{sec:galaxies}). 
Next, we apply this methodology to the study of the environment of a $z\approx 3.5$ Lyman Limit System (LLS) in the line of sight to the quasar \qname\ (Sect.~\ref{sec:thisdla}). This system is selected for our initial analysis for its particulaly interesting chemical composition, being one of the very few systems currently known of extremely metal-poor gas clouds with chemical properties that are consistent with the remnants of Population III stars \citep{crighton2016}. We conclude  with a summary and an outlook of future studies that this survey will enable (Sect.~\ref{sec:summary}). Through this and subsequent work, unless otherwise specified, we make use of a Planck 15 cosmology 
\citep[ $\Omega_m = 0.307$, $H_0 = 67.7~\rm km~s^{-1}~Mpc^{-1}$;][]{planck2016}, we assume the AB magnitude system, and we express distances in proper (physical) units. 

\afterpage{
\begin{landscape}
\begin{table}
\centering
\caption{Summary of the sample properties and MUSE observations.
The table lists: the reference name of the quasar; the common name in the NASA/IPAC Extragalactic Database (NED);
the right ascension and declination of the quasar at the centre of the field (J2000); the quasar $r$ band magnitude 
with its associated uncertainty; the quasar redshift derived
from rest-frame UV lines with its associated uncertainty; the number of known strong absorption line systems with $N_{\rm HI} \gtrsim 10^{17}~\rm cm^{-2}$; the total on-source exposure time of MUSE observations; the programme number under which observations have been collected; the resulting image quality in the
reconstructed $r$ band from MUSE data; the limiting flux (1$\sigma$ pixel rms) measured at $5500~\rm \AA$ and the corresponding AB magnitude (2$\sigma$) for a 0.7~arcsec aperture.}\label{tab:sample}
\begin{tabular}{llllcccclcc} 
\hline
Name & Common Name & R.A. & Dec. & $m_{\rm r}$ & $z_{\rm qso,uv}$ & $N_{\rm lls}$ &  $t^{\rm MUSE}_{\rm exp}$ & PID & I.Q. & $F_{\rm rms}/m_{\rm ap}$ \\
 &  & (hh:mm:ss) & (dd:mm:ss) & (mag) & &  & (hours) &  & ($\rm arcsec$)  & ($10^{-20}\rm erg~s^{-1}~cm^{-2}~\AA^{-1}$/mag) \\

\hline
J010619.24+004823.3$^{a}$  & SDSS J010619.24+004823.3     & 01:06:19.24 & +00:48:23.31   & 19.10(0.01) & 4.4402(0.0002)     & 2 & 4.02 & 197.A$-$0384               & 0.67 &1.84/26.25 \\
J012403.77+004432.7$^{a}$  & SDSS J0124+0044	          & 01:24:03.77 & +00:44:32.76   & 17.95(0.01) & 3.8359(0.0003)     & 2 & 4.12 & 197.A$-$0384,096.A$-$0222  & 0.73 &1.99/26.16 \\
J013340.31+040059.7$^{a}$  & SDSS J013340.31+040059.7     & 01:33:40.31 & +04:00:59.77   & 18.45(0.01) & 4.1709(0.0002)     & 3 & 4.02 & 197.A$-$0384               & 0.63 &1.86/26.24 \\
J013724.36$-$422417.3$^{c}$& BRI J0137$-$4224             & 01:37:24.36 & $-$42:24:17.30 & 18.46(0.05) & 3.975(0.012)$^d$   & 2 & 4.82 & 197.A$-$0384               & 0.68 &1.41/26.54 \\
J015741.56$-$010629.6$^a$  & SDSS J015741.56$-$010629.5   & 01:57:41.56 & $-$01:06:29.66 & 18.30(0.01) & 3.5645(0.0001)     & 2 & 4.02 & 197.A$-$0384               & 0.77 &1.78/26.29 \\
J020944.61+051713.6$^a$    & SDSS J020944.61+051713.6     & 02:09:44.61 & +05:17:13.66   & 18.44(0.01) & 4.1846(0.0006)     & 3 & 4.55 & 197.A$-$0384               & 0.57 &1.58/26.41 \\
J024401.84$-$013403.7$^a$  & BRI 0241$-$0146	          & 02:44:01.84 & $-$01:34:03.78 & 18.18(0.01) & 4.044(0.012)$^d$   & 2 & 4.02 & 197.A$-$0384               & 0.67 &1.62/26.39 \\
J033413.42$-$161205.4$^b$  & BR 0331$-$1622	          & 03:34:13.42 & $-$16:12:05.36 & 18.63(0.01) & 4.380(0.013)$^d$   & 2 & 4.02 & 197.A$-$0384               & 0.67 &1.66/26.36 \\
J033900.98$-$013317.7$^c$  & PKS 0336$-$017	          & 03:39:00.98 & $-$01:33:17.70 & 19.17(0.05) & 3.204(0.009)$^d$   & 2 & 4.02 & 197.A$-$0384               & 0.62 &1.61/26.40 \\
J094932.26+033531.7$^a$    & SDSS J094932.26+033531.7     & 09:49:32.26 & +03:35:31.78   & 18.03(0.01) & 4.1072(0.0004)     & 2 & 4.02 & 197.A$-$0384               & 0.69 &1.82/26.26 \\
J095852.19+120245.0$^a$    & [HB89] 0956+122	          & 09:58:52.19 & +12:02:45.04   & 17.47(0.01) & 3.2746(0.0003)     & 2 & 4.10 & 094.A$-$0280               & 0.60 &2.95/25.73 \\
J102009.99+104002.7$^a$    & [HB89] 1017+109	          & 10:20:09.99 & +10:40:02.73   & 17.72(0.01) & 3.1528(0.0003)     & 1 & 4.40 & 096.A$-$0937               & 0.69 &2.07/26.12 \\
J111008.61+024458.0$^a$    & SDSS J111008.61+024458.0     & 11:10:08.61 & +02:44:58.07   & 18.28(0.01) & 4.1582(0.0003)     & 2 & 4.02 & 197.A$-$0384               & 0.72 &1.97/26.17 \\
J111113.79$-$080402.0$^b$  & BRI 1108$-$0747	          & 11:11:13.79 & $-$08:04:02.00 & 18.49(0.01) & 3.930(0.012)$^d$   & 3 & 4.41 & 197.A$-$0384,095.A$-$0200  & 0.70 &2.26/26.02 \\
J120917.93+113830.3$^a$    & [HB89] 1206+119	          & 12:09:17.93 & +11:38:30.34   & 17.45(0.01) & 3.0836(0.0001)     & 2 & 3.96 & 197.A$-$0384,094.A$-$0585  & 0.66 &1.88/26.22 \\
J123055.57$-$113909.3$^c$  & BZQ J1230$-$1139	          & 12:30:55.57 & $-$11:39:09.30 & 19.84(0.05) & 3.557(0.012)$^d$   & 1 & 4.02 & 197.A$-$0384               & 0.66 &1.75/26.30 \\
J124957.23$-$015928.8$^a$  & SDSS J124957.23$-$015928.8   & 12:49:57.23 & $-$01:59:28.80 & 17.78(0.01) & 3.6337(0.0003)     & 1 & 4.02 & 197.A$-$0384               & 0.65 &2.10/26.11 \\
J133254.51+005250.6$^a$    & SDSS J133254.51+005250.6     & 13:32:54.51 & +00:52:50.63   & 18.35(0.01) & 3.5071(0.0001)     & 1 & 4.02 & 197.A$-$0384               & 0.65 &1.62/26.38 \\
J142438.10+225600.7$^a$    & RX J1424.6+2255	          & 14:24:38.10 & +22:56:00.71   & 15.17(0.01) & 3.634(0.012)$^d$   & 1 & 4.00 & 095.A$-$0200,099.A$-$0159  & 0.83 &2.29/26.01 \\
J162116.92$-$004250.8$^a$  & SDSS J1621$-$0042	          & 16:21:16.92 & $-$00:42:50.86 & 17.28(0.01) & 3.7100(0.0002)     & 1 & 9.76 & 095.A$-$0200,097.A$-$0089  & 0.65 &1.92/26.20 \\
J193957.25$-$100241.5$^c$  & PKS1937$-$101 	          & 19:39:57.25 & $-$10:02:41.50 & 16.61(0.05) & 3.787(-)$^e$       & 1 & 4.61 & 197.A$-$0384,094.A$-$0131  & 0.80 &2.00/26.16 \\
J200324.14$-$325144.8$^c$  & [HB89] 2000$-$330	          & 20:03:24.14 & $-$32:51:44.80 & 17.38(0.05) & 3.785(0.011)$^d$   & 3 & 10.00& 094.A$-$0131               & 0.74 &1.45/26.51 \\
J205344.72$-$354655.2$^c$  & [WHO91] 2050$-$359	          & 20:53:44.72 & $-$35:46:55.20 & 18.41(0.05) & 3.490(-)$^e$       & 2 & 4.02 & 197.A$-$0384               & 0.66 &2.16/26.07 \\
J221527.29$-$161133.0$^b$  & BR 2212$-$1626	          & 22:15:27.29 & $-$16:11:33.00 & 18.13(0.01) & 4.000(0.013)$^d$   & 2 & 4.02 & 197.A$-$0384               & 0.68 &2.26/26.02 \\
J230301.45$-$093930.7$^a$  & SDSS J230301.45$-$093930.6   & 23:03:01.45 & $-$09:39:30.72 & 17.68(0.01) & 3.4774(0.0003)     & 1 & 4.02 & 197.A$-$0384               & 0.69 &1.75/26.30 \\
J231543.56+145606.4$^{a}$  & SDSS J231543.56+145606.3     & 23:15:43.56 & +14:56:06.41   & 18.54(0.01) & 3.3971(0.0004)     & 2 & 4.28 & 197.A$-$0384               & 0.73 &1.77/26.29 \\
J233446.40$-$090812.2$^a$  & FBQS J2334$-$0908	          & 23:34:46.40 & $-$09:08:12.24 & 18.03(0.01) & 3.3261(0.0005)     & 2 & 4.02 & 197.A$-$0384               & 0.74 &1.66/26.36 \\
J234913.75$-$371259.2$^b$  & BR J2349$-$3712	          & 23:49:13.75 & $-$37:12:59.25 & 19.15(0.02) & 4.240(0.012)$^d$   & 2 & 4.28 & 197.A$-$0384               & 0.71 &1.56/26.43 \\
\hline
\end{tabular}
\flushleft{$^a$ Entries related to the quasar (R.A., Dec., $m_{\rm r}$ and $z_{\rm qso,uv}$) are, unless otherwise noted, from SDSS \citep{abolfathi2017}.
$^b$ Quasar photometry (R.A., Dec., $m_{\rm r}$) is from the ATLAS survey \citep{shanks2015}. $^c$ Quasar photometry (R.A., Dec., $m_{\rm r}$) is from the Million Quasars Catalog \citep{flesch2015,flesch2017}. $^d$ Redshift from this work (see text). $^e$ Redshift from NED.}
\end{table}
\end{landscape}
}

\section{Sample Selection and Survey Strategy}\label{sec:selection}

Our survey is designed to investigate the connection between optically-thick gas and galaxies at $z\approx 3-4$. For this purpose, we selected a sample of quasars at $z\gtrsim 3.2$ for which high-resolution ($R\gtrsim 30,000$) spectroscopy was available (as of 2014) from the Ultraviolet and Visual Echelle Spectrograph  \citep[UVES;][]{uves2000} at VLT, the Magellan Inamori Kyocera Echelle \citep[MIKE;][]{mike2003} at Magellan, and the High Resolution Echelle Spectrometer \citep[HIRES;][]{hires1994} at Keck.
This results in a sample of quasars with  magnitudes  $m_{\rm r} \lesssim 19~\rm mag$. We further restrict our sample to quasars with data at moderate or high signal-to-noise
($S/N \gtrsim 20$), and with at least one strong absorption line system ($N_{\rm HI}\gtrsim 10^{17}~\rm cm^{-2}$) at redshift $z\gtrsim 3.05$ (the lowest Ly$\alpha$ redshift accessible at the bluest wavelengths of MUSE).
Finally, we restrict the sample to quasars that are observable from Paranal with low airmass, typically at declination $\delta < +15~\rm deg$.

Our final selection comprises \nqso\ quasars (Table~\ref{tab:sample}), including archival sightlines that have been observed as part of the
guaranteed time observations (GTO) by the MUSE consortium \citep[e.g.][]{borisova2016} and the sightline presented in \citet{fumagalli2016}. In the end, our sample contains \nlls\ strong absorption line systems, including damped Ly$\alpha$ absorbers (DLAs). This sample does not have a particular selection function, and it is assumed to be representative of the general population of absorbers at these redshifts particularly because no pre-selection has been made regarding, e.g., the chemical composition or the kinematics of the absorbers (see also Sect.~\ref{sec:als}).

As part of the programme ID 197.A-0384, we have observed each quasar field with 5 observing blocks (OBs) of 1 hour with MUSE between period
97 and period 103. After excluding overheads, this corresponds to a total on-source observing time of $\approx 4$ hours per field, with longer
exposure times in GTO fields (up to 10 hours) or fields with partial MUSE observations from the archive (\citealt{arrigoni2019}, see Table~\ref{tab:sample}).
Each OB is structured in a sequence of $3\times 960~\rm s$ exposures, with relative rotations of 90 degrees and small dithers
($\approx 1~\rm arcsec$) that are designed to mitigate spatial inhomogeneities in the data that arise from the small differences
in the performance of the MUSE spectrographs. Observations are completed in service mode, and thus are only executed on clear nights at airmass $\lesssim 1.6$ when the image quality is of the order of $0.8~\rm arcsec$ or better. Occasionally, for our data and in some archival data, the resulting image quality is above our requirements, but not in excess of $\approx 0.9~\rm arcsec$.
With this observing strategy, our programme is designed to deliver a homogeneous spectroscopic survey to a flux  limit of $\sim 4\times 10^{-18}~\rm  erg~s^{-1}~cm^{-2}~\AA^{-1}$ ($S/N > 5$ at $\lambda = 5550~\rm \AA$, see Fig.~\ref{fig:LAEcompleteness}) in a region of $\approx 500\times 500~\rm kpc^2$ at $z\approx 3$, which is centred at the quasar position. 

\section{Data Reduction}\label{sec:reduction}

\begin{table*}
\centering
\caption{Summary of the archival quasar spectroscopy used in this survey.
The first three entries are shown, and the full table is included as online only material.
  The table lists: the quasar name; the instrument name; the spectral resolution (range is used if dependent on instrument arm);
  the wavelength range covered by the spectrum (gaps may be present); the typical $S/N$ per pixel representative of the Ly$\alpha$ forest, away from saturated absorption lines (actual wavelengths given in parenthesis, in \AA);
the typical $S/N$ per pixel measured representative of the continuum redward to the quasar Ly$\alpha$ (actual wavelengths given in parenthesis, in \AA); the nominal pixel velocity dispersion of the 1D spectra (range is used if arm dependent).}\label{tab:qsospectra}
\begin{tabular}{llccrrr} 
\hline
Name & Instrument & Resolution          & Wavelength Range & $S/N_{\rm blue}$  & $S/N_{\rm red}$ & Dispersion      \\
     &            &                     &       (\AA)      &               &              &    ($\rm km~s^{-1}$) \\

\hline
\hline
J010619.24+004823.3      &  MIKE & 28,000;22,000 & 3450$-$8100 &  7 (6280)  & 10 (7000)  &    4.2 \\
                         &  ESI  & 5,400           & 3927$-$11068&  31 (6280) & 36 (7000)  &    10  \\
\hline
J012403.77+004432.7      &  UVES & 40,000 & 4173$-$6813 & 27 (5500) & 23 (6500)  &    2.5  \\	 
                         &  X-SHOOTER & 4350;7450;5300 & 3150$-$24800 & 31 (5500) & 33 (6500)  & 20;11;19   \\	 
\hline
J013340.31+040059.7      &  UVES & 40,000 & 4665$-$10425& 20 (5500) & 15 (7000)  &    2.5  \\	 
                         &  HIRES& 49,000 & 4160$-$8720 & 14 (5500) & 15 (7000)  &    1.3  \\	 
                         &  ESI  & 5,400 & 3927$-$11068& 25 (5500) & 30 (7000)  &  10  \\
                         &  X-SHOOTER & 4350;7450;5300 & 3150$-$18000 & 55 (5500) & 54 (7000)  & 20;11;19   \\	 
\hline
\end{tabular}
\end{table*}

\subsection{Archival quasar spectroscopy}
High dispersion spectroscopy of the quasars targeted in this survey are collected from the VLT and the Keck archives, and it is also supplemented with data acquired at the Magellan telescopes at Las Campanas Observatory.
We further complement this data set with moderate dispersion spectroscopy from the  Echellette Spectrograph and Imager \citep[ESI;][]{esi2002} at Keck and X-SHOOTER \citep{xshooter2011} at the VLT.
Details on data available for each quasar sightline are listed in Table~\ref{tab:qsospectra}, where we summarise the main observational information, including: the representative wavelength range covered by the data, the spectral resolution, and the final $S/N$ ratio at selected wavelengths.
In the following, we briefly describe the data processing for each instrument.

The HIRES data are drawn from the Keck Observatory Database of Ionized Absorption toward Quasars (KODIAQ) DR1 and DR2 sample of quasars \citep{lehner2014, omeara2015,omeara2017}. A detailed description of the data reduction and continuum fitting procedure is in \citet{omeara2015}. Briefly, sets of observations collected with the same instrument setup are reduced with the {\sc HIRedux} pipeline\footnote{\url{http://www.ucolick.org/~xavier/HIRedux/}.} that performs basic processing (bias subtraction, flat-fielding), and determines a wavelength solution (using vacuum wavelengths) for the chips. After sky subtraction, the objects are  extracted on an order-by-order basis and, if multiple exposures are present, data are weighted-mean combined. Following this step, data are continuum normalized using Legendre polynomial fits to each spectral order. In the Ly$\alpha$ forest and blueward to that, the continuum is determined at reference points judged to be absorption free. For this work, if  multiple observations of a quasar exist (e.g. from different observers, or with differing HIRES setups), we further combine the data into a single spectrum.  The combination is performed by resampling the spectrum onto a common wavelength solution, and summing the spectra, weighting by their $S/N$.

The ESI data for this program are drawn from the KODIAQ DR3 sample of quasars (O'Meara et al. 2019, in prep.). Full details of the data reduction for these quasars will appear in a forthcoming publication, but a short summary is as follows. The raw data are downloaded from the Keck Observatory Archive (KOA) and organized by observing run.  The data span the date range 2000-2014.  ESI is a fixed wavelength range instrument, so only the binning and slit width can change from observation to observation.  For each observing run, the data were processed using the {\sc esi\_redux} package\footnote{\url{https://www2.keck.hawaii.edu/inst/esi/ESIRedux/}.}.  The pipeline subtracts the bias, applies flat fields, and corrects scattered light on each frame, then optimally extracts each object and places it on a wavelength scale derived from arc lamp observations.  Multiple observations of the same object within an observing run with the same slit and binning are combined, weighted by their $S/N$.  The data are then flux calibrated using spectrophotometric standard star observations, and the echelle orders are then combined into a single one dimensional (1D) flux spectrum with associated errors. For this work, we choose the highest $S/N$ spectrum for analysis. For objects with multiple spectra of comparable $S/N$, we choose the spectrum with the highest resolution as determined by the slit width. A continuum model is further derived for each spectrum, to enable the analysis of absorption line systems, using Legendre polynomial fits to the data or spline points blueward of the quasar Ly$\alpha$ emission line, selected in regions that are deemed by visual inspection to be free from absorption. 

The UVES spectra are processed as part of the UVES Spectral Quasar Absorption Database (SQUAD) project \citep{murphy2019}. The SQUAD data reduction procedure utilises the standard ESO pipeline for UVES (version 4.7.8) with the improved wavelength calibration line list and procedures described in \citet{murphy2007}. This pipeline bias- and flat-field corrects the raw UVES images, defines the echelle order numbers and locations for each exposure using a physical model of the spectrograph and dedicated short-slit flat-field and thorium-argon exposures, optimally extracts and blaze-corrects the quasar exposure, and attaches a wavelength solution derived from a thorium-argon exposure with matching spectrograph settings (including slit-width). In the SQUAD, the thorium-argon signal is extracted using the same object weights used to extract the corresponding quasar exposure.

The extracted spectra from all orders and all exposures are then combined into a final spectrum with associated error array for each quasar, using the custom-written code {\sc uves\_popler}\footnote{POst PipeLine Echelle Reduction software, Murphy. M. 2016, \url{doi:10.5281/zenodo.44765}.}. Details of this process are described in \citet{murphy2016}. Briefly, {\sc uves\_popler} re-disperses all spectra onto a common (vacuum-heliocentric) wavelength grid with $2.5~\rm km~s^{-1}$ pixels (all quasar exposures had 2$\times$2 on-chip binning), automatically removes some artefacts (e.g.\ cosmic rays, stray internal reflections, poorly-extracted data, residual blaze correction errors), and allows manual identification and removal of any remaining artefacts. Next, the code scales the spectra for optimal combination via a $\sigma$-clipped weighted mean, and sets an initial continuum using polynomial fits to small, overlapping portions of the spectrum. The continuum redwards of the Ly$\alpha$ emission needed little manual adjustment, except near some broader absorption features. However, because of the large number of absorption lines, the continuum in the Ly$\alpha$ forest region is entirely re-estimated manually using polynomial fits.  For most quasars, all the exposures were taken with the same slit width. However, in 3 cases (J013340$+$040059, J111113$-$080402, J233446$-$090812), a minority of exposures has a slightly ($0.1-0.2$ arcsec) narrower slit.
Due to the adopted combination algorithm, Table~\ref{tab:qsospectra} provides only a representative slit width.
The representative slit width is used to provide a nominal resolving power, $R$, assuming an $R$ to slit-width ratio of 40000 which is a compromise between the slightly higher value for the blue arm of UVES ($R=41400$) and the lower value for the red arm ($R=38700$).

The X-SHOOTER data are retrieved from the phase 3 release of the XQ-100 survey \citep{lopez2016}. The only exceptions are the observations of  J015741$-$010629 and J020944$+$051713, which are retrieved from the ESO archive, but have been processed in the same way as the XQ-100 data. Details on the observations and data reduction are presented in \citet{lopez2016}\footnote{See also \url{http://www.eso.org/sci/observing/phase3/data_releases/xq100_dr1.pdf}.}. Briefly, data are reduced using a custom {\sc IDL} pipeline (developed by G. Becker), designed to obtain a better removal of the background and to perform an optimal extraction of the spectra.
Individual frames are bias or (for the NIR arm) dark subtracted, and flat-fielded.
After sky subtraction, the two-dimensional frames are flux calibrated using observations of spectro-photometric standard stars. A single 1D spectrum is then extracted from each exposure of a given object in each arm. Data are then re-sampled on a uniform wavelength grid (vacuum-heliocentric system) in each arm. An additional spectrum is also produced by joining the spectra of the three arms together. Telluric absorption in the VIS and NIR arms have been corrected for by modelling the one-dimensional spectra separately using {\sc Molecfit} \citep{smette2015}. 
Finally, a continuum model is derived for each arm by selecting points along the quasar continuum in regions free of absorption as knots for a cubic spline fit.

\begin{figure*}
\begin{tabular}{l}
\includegraphics[scale=0.5]{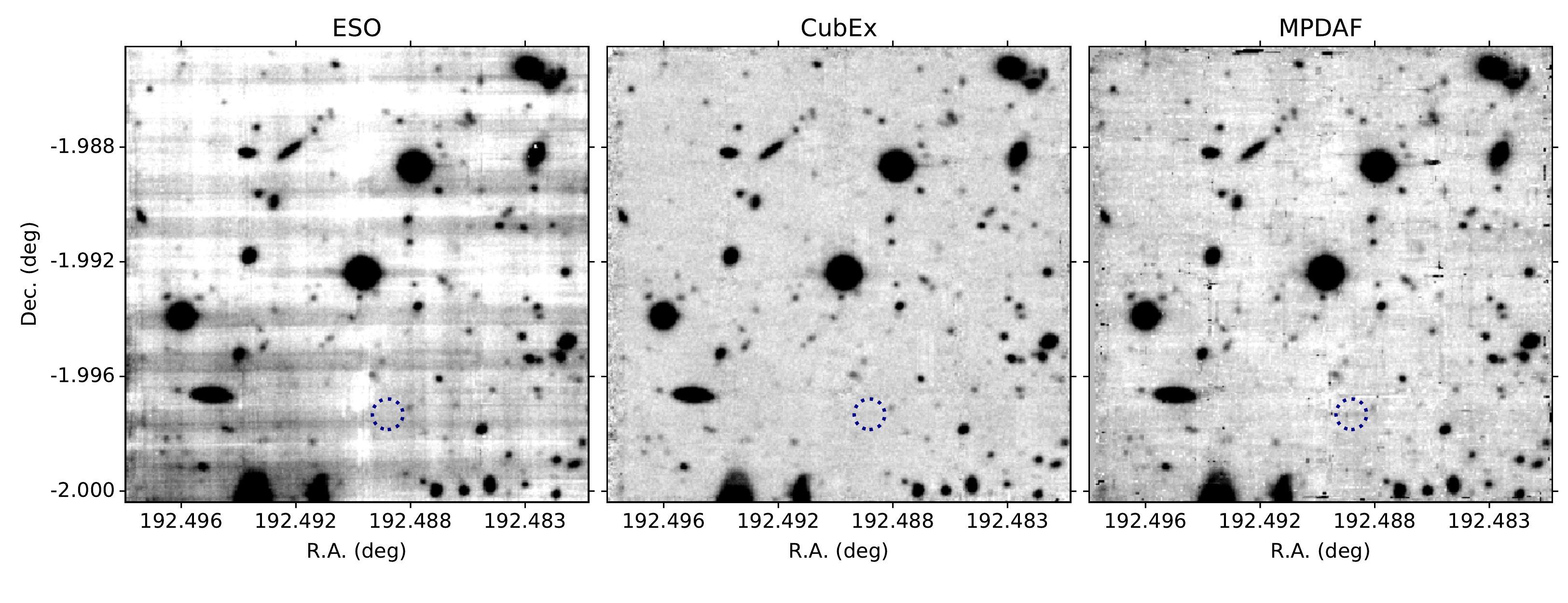}\\
\includegraphics[scale=0.52]{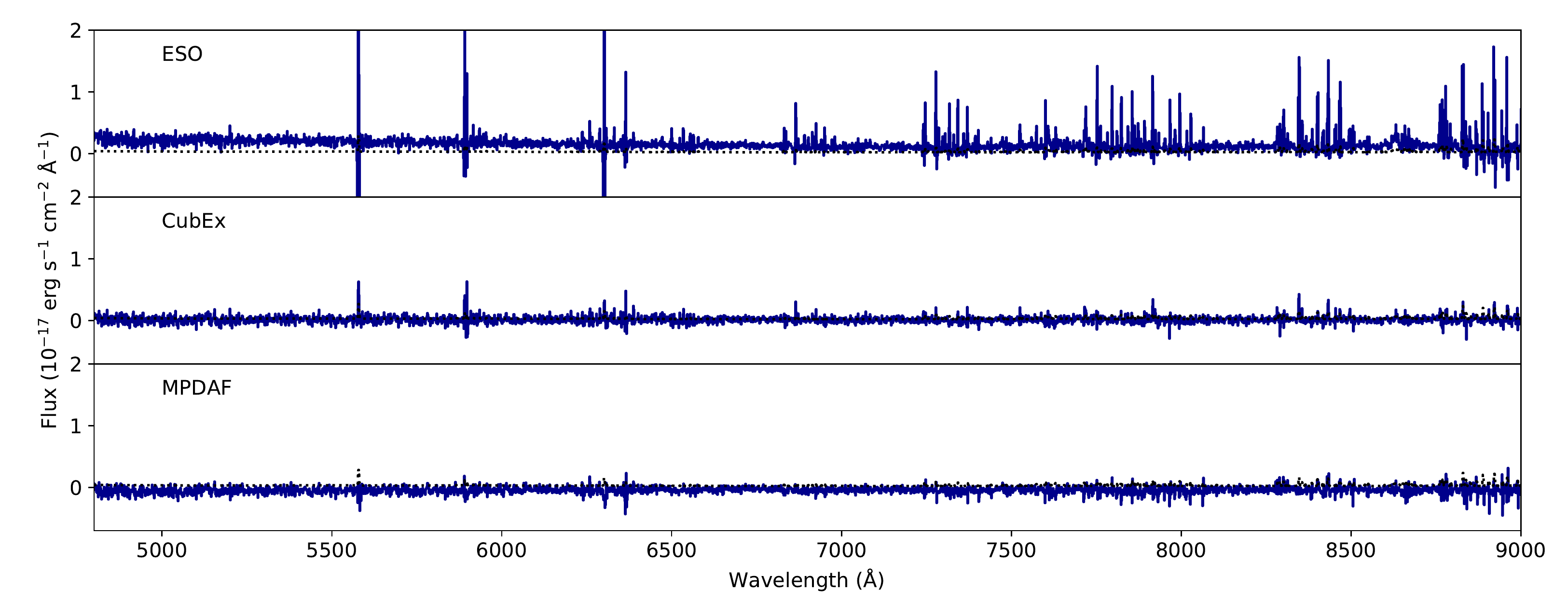}
\end{tabular}
\caption{Comparison of the final products of the three reduction procedures adopted in this work in the field \qname, which we analyse in detail this paper.  ({\it Top}) White-light image reconstructed by collapsing the final datacubes along the wavelength axis for different reduction pipelines (as labelled). 
The three panels are on the same flux scale, to highlight how different reduction techniques correct the residual illumination fluctuations across IFUs and slices.  ({\it Bottom}) Spectrum of an empty sky region from each of the final datacubes (as labelled), showing the performance of the different sky subtraction algorithms, especially for sky lines at  $\lambda>7000$~\AA~ and in the continuum at $\lambda<5500$~\AA. The post-processing techniques employed in this analysis successfully reduce systematic fluctuations in the original data.}  \label{fig:museredux}
\end{figure*}

\subsection{MUSE Spectroscopy}

The reduction of MUSE data follows a multi-step process, as detailed below. 
In particular, the raw data are initially processed using the ESO Muse pipeline \citep{weilbacher2014} and after this step each individual exposure is post-processed with two independent packages: CubExtractor (hereafter CubEx; Cantalupo, in prep., see Cantalupo et al. 2019 for a description) and MPDAF \citep{piqueras2017}, to improve the quality of the final datacubes. This software has been widely used in the community in the last four years (for some recent examples see, e.g.: \citet{marino2018, cantalupo2019, feruglio2019, lusso2019, mackenzie2019, nanayakkara2019}.

\subsubsection{ESO pipeline reduction}

The first part of our reduction pipeline is based on the recipes distributed as part of the ESO MUSE pipeline \citep[][version 2 or greater]{weilbacher2014}, which processes the raw data and applies standard calibrations to the science exposures.
Briefly, the pipeline generates a master bias, a master flat, processes the arcs, and reduces the sky flats.
Next, calibrations are applied to the standard star and a sensitivity function is then generated. Finally, these calibrations are applied to the raw science exposures and data cubes with associated pixel tables are reconstructed.

At this stage, we also reconstruct cubes that are sky subtracted by the ESO pipeline using models of the sky continuum and sky lines that are
computed using the darkest pixels in the FOV. After aligning the individual exposures by using point sources in the field,
we generate a stack of all science frames into a single final cube, which we dub the ``ESO product''. Finally, we register this final stack on a reference coordinate system by imposing an absolute zero point for the World Coordinate System using the position of the quasar at the centre of the field. For our reference system, we use Gaia astrometry \citep{gaia2018}.

The final ESO stack is known to have some imperfections arising both from second-order variations in the illumination of the
detectors, and residuals associated to the subtraction of sky lines \citep{bacon2017}. Tools to mitigate these imperfections have
been developed and are employed in this survey as described below. 
For this reason, we do not use the ESO cube for science, but employ this data product as a
reference grid for further post-processing using {\sc CubExtractor} ({\sc CubEx} hereafter)  and {\sc MPDAF}.

\subsubsection{The {\sc CubEx} pipeline reduction}\label{sec:cubexredux}

Following the standard reduction using the ESO pipeline, we post-process individual exposures using the tools distributed as part of {\sc CubEx} (v1.8). 
In the following, we briefly describe the algorithms and the adopted procedure (for a more detailed description of these algorithms, see, e.g., Cantalupo et al. 2019).

The first step is to reconstruct a data cube for each exposure resampling the pixel tables onto the reference grid defined
by the final cube generated using the ESO pipeline, as described above. For this step, we use the {\sc muse\_scipost} recipe within the ESO pipeline. As we are starting from the pixel table, this is the only step in which we resample the data onto a regular grid. The subsequent operations are performed on the reconstructed data cubes. 

Next, we use the {\sc CubeFix} tool to correct residual differences in the relative illumination of the 24 MUSE IFUs and of individual slices,
which are not completely corrected by the flat-fields. {\sc CubeFix} scans the cube as a function of wavelength to re-align the relative illumination
of IFUs and slices, and further adjusts the relative illumination of ``stacks'' (each MUSE IFU is composed of 4 stacks of 12 slices) with
white-light images reconstructed from the cube. After this step, we use the {\sc CubeSharp} tool for sky subtraction. 
{\sc CubeSharp} implements an algorithm to perform local sky subtraction, including empirical corrections of the sky line spread function
(LSF).  This step, which is flux-conserving, enables a more accurate removal of the sky lines compared to the ESO pipeline reduction, minimizing residuals
that arise from variation in the line spread function across the MUSE IFUs.
The combination of {\sc CubeFix} plus {\sc CubeSharp} post-processing is applied twice, by using the first illumination-corrected and sky-subtracted cube to mask continuum-detected sources during the second iteration of {\sc CubeFix} and {\sc CubeSharp}, thus enabling a more accurate determination of the mean illumination and background of each slice. 

After this step, we combine all the individual exposures in a single data cube, using an average $3\sigma$ clipping algorithm. Edges of individual IFUs are masked at this stage, and individual exposures are inspected to manually mask any residual artefacts via a custom graphical user interface. 
From this high $S/N$ datacube, we create a white-light image from which we identify continuum sources and create an updated source mask. We then input this back into a final iteration of corrections with {\sc CubeFix} and {\sc CubeSharp}, which reduces any contamination from the identified sources in the illumination corrections. 
In the end, we reconstruct four final data products,
including an average cube of all exposures (mean cube), a median cube, and two cubes  (combined with both mean and median statistics) containing only one half of all the exposures each, which are useful to identify contaminants such as residual cosmic rays. 
Figure~\ref{fig:museredux} shows a comparison between the final products processed with the ESO and the {\sc CubEx} pipelines, highlighting
the relative improvement over the basic pipeline reduction.

\subsubsection{The {\sc MPDAF} pipeline reduction}
The preparation of the third data product follows a similar procedure to the one described in \citet{fumagalli2017uvb}, but relies on the self-calibration method described in \citet{bacon2017} and included in the {\sc MPDAF} package\footnote{This method is implemented in the standard MUSE ESO pipeline from version 2.4 onward.} \citep[][v3.0]{piqueras2017}

As a first step, individual exposures are re-sampled on a common astrometric grid defined by the final ESO product. This is the only step in which data are resampled. Next, residual imperfections in the flat-fielding are corrected using the self-calibration tool implemented in {\sc MPDAF}. This procedure implements a similar algorithm to the one in {\sc CubeFix}, i.e. it re-aligns the flux scale in each slice as a function of wavelength, but it operates directly on the pixel tables rather than on the reconstructed cubes. 
After reconstructing data cubes using the ESO {\sc muse\_scipost\_make\_cube} recipe, we perform sky subtraction using the Zurich Atmosphere Purge ({\sc ZAP}) code \citep[]{soto2016}, which performs a principal component analysis of the sky to separate sky-lines from continuum sources. For both steps, we mask bright sources using a white-light image reconstructed from the {\sc CubEx} final datacubes. The sky-subtracted cubes are finally combined in a single datacube using both mean and median statistics.

\subsubsection{Quality assurance}

Fig.~\ref{fig:museredux} compares the final products derived with the three pipelines. Both the {\sc CubEx} and the {\sc MPDAF} reduction procedures are successful in improving the sky subtraction, especially for sky lines at  $\lambda>7000$~\AA~ and in the continuum at $\lambda<5500$~\AA. To quantify the relative improvement with respect to the ESO reduction, we compute the ratio of the flux standard deviation in a sky spectrum within the range 6000-9000~\AA, finding 0.30 for the {\sc CubEx} reduction compared to the ESO one, and 0.32 for the  {\sc MPDAF} reduction relative to ESO. 

Likewise, both the {\sc CubEx} and the {\sc MPDAF} reduction pipelines significantly improve the quality of the illumination homogeneity across different IFUs and slices, with the {\sc CubEx} product achieving the best result in terms of flux homogeneity for our observational strategy. The uniformity of the illumination can be quantified by comparing the flux standard deviation of sky pixels in the white-light image from the {\sc CubEx} and {\sc MPDAF} processing relative to the ESO reduction. We find ratios of 0.14 and 0.31, respectively, confirming the visual impression from Fig.~\ref{fig:museredux}. For this reason, throughout this survey, we will use primarily the products of the {\sc CubEx} reduction for our analysis. We retain the MPDAF products, however, so that they can be used as an extra check on our results to ensure that any identified features are real and not an artefact of the post-processing steps. 

\begin{figure}
\includegraphics[scale=0.55]{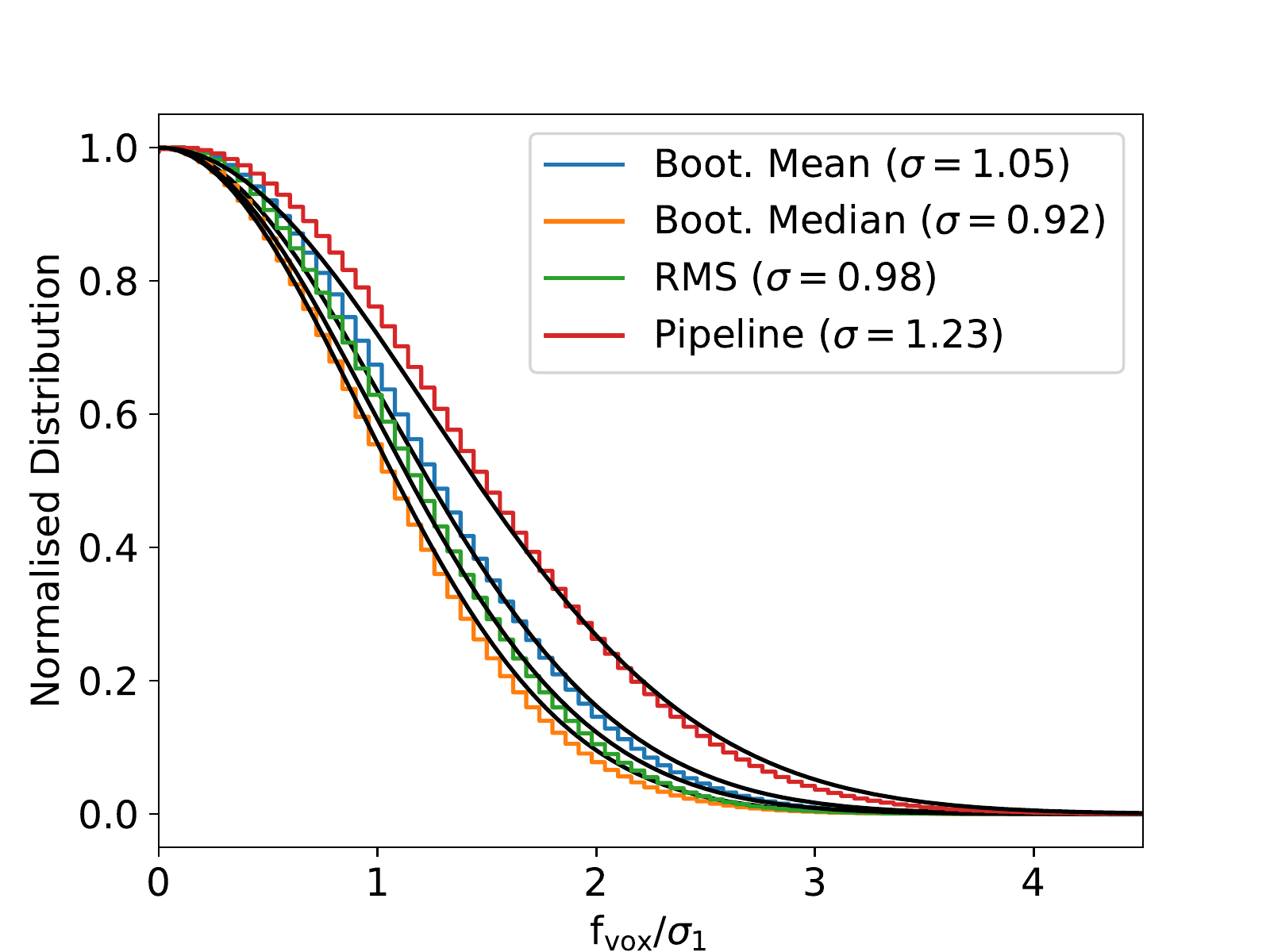}
\caption{Histograms of flux in voxels ($f_{\rm vox}$) in the range $4900-5500~\rm \AA$, normalised by the propagated standard deviation in the same voxel ($\sigma_1$) from the ESO pipeline (red), after rescaling $\sigma_1$ to match the voxel standard deviation (green), and from bootstrap resampling (blue and yellow for the mean and median, respectively). Shown with black lines are Gaussian fits to these distributions, with the resulting standard deviation listed in the legend. Voxels overlapping with continuum sources have been excluded from this analysis. If the noise is correctly normalised, this distribution should approximate a Gaussian with standard deviation of unity. The rescaled variance adopted throughout our analysis is a better representation of the true noise compared to the original pipeline values.}\label{fig:rmshisto}
\end{figure}

To validate the photometric calibration of our fields, we compare the $r$-band aperture magnitudes obtained from the datacubes against the Petrosian magnitude from SDSS. We find that for 443 bright sources ($m_r<22$~mag), the median difference between the MUSE magnitudes and the SDSS ones is less than 3 per cent. 
Similarly, we compare the quasar spectra extracted from the MUSE cubes with the archival spectroscopy described above, finding excellent agreement with respect to wavelength calibration. 
Finally, we measure the resulting image quality on the reconstructed $r-$band images by fitting a two-dimensional Moffat function to point sources in the fields. The resulting full widths at half maximum are listed in Table~\ref{tab:sample}, showing that we achieve the desired image quality ($\lesssim 0.8$~arcsec) for all the fields. In this table, we also list the 1$\sigma$ root-mean-square (rms) of background pixels computed in a 50~\AA\ window centred at 5500~\AA\ in each cube, as a metric of the achieved depth in our observations.  
These values are also converted to 2$\sigma$ AB limiting magnitudes assuming a 0.7~arcsec aperture.

\begin{figure}
\includegraphics[scale=0.55]{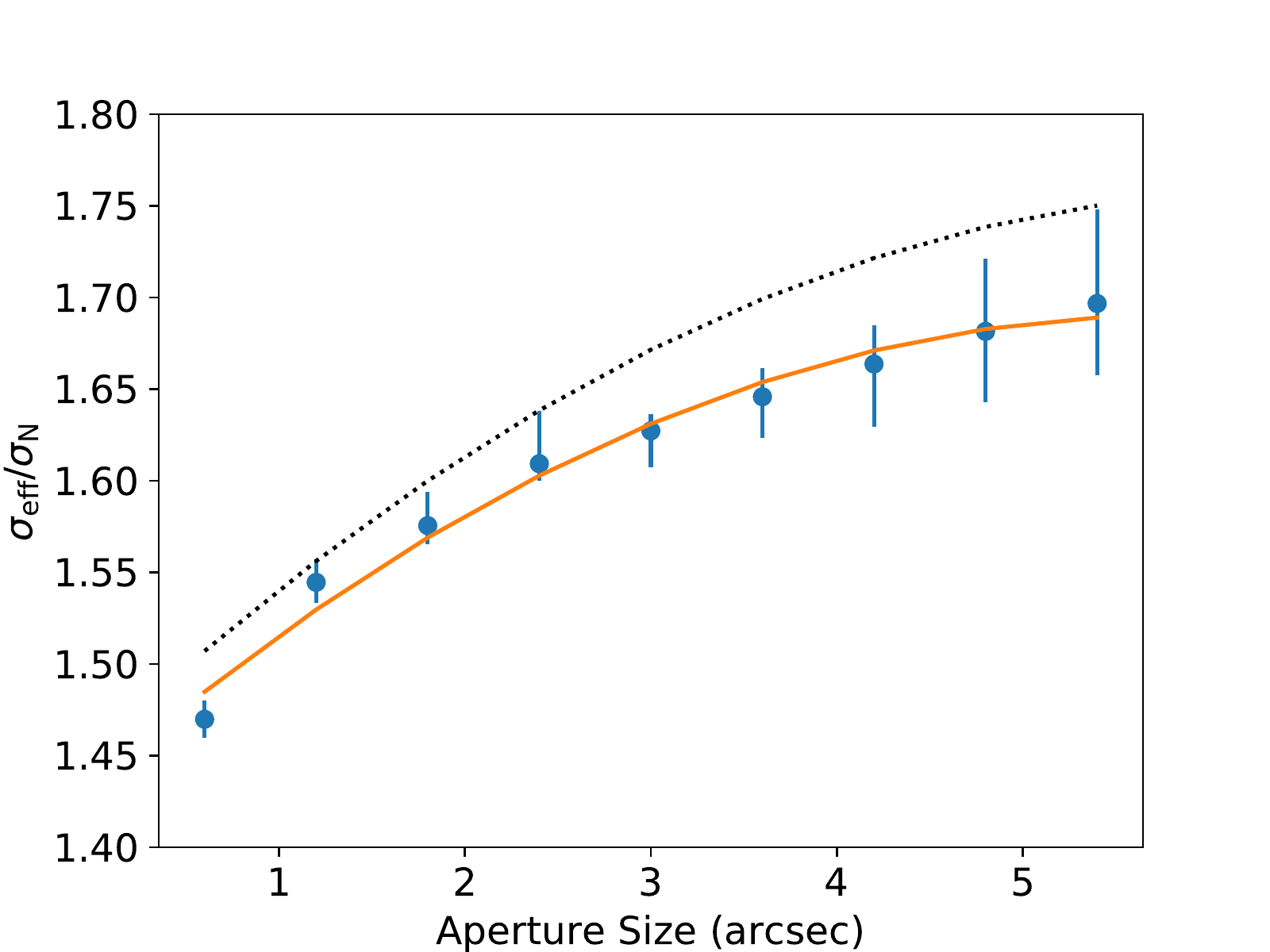}
\caption{Median (blue points) and interquartile range (bars) of the ratio between the flux dispersion in apertures of varying size ($\sigma_{\rm eff}$) and the error computed propagating the variance ($\sigma_{\rm N}$), for all the MUSE cubes in a $5~\rm \AA$ window (4 pixels) between $4900-5500~\rm \AA$. A second-order polynomial fit (orange) provides a correction function that can be used to estimate the effective $S/N$ of the detections within the data cubes. The dotted line represents the same correction for a spectral window of $7.5~\rm \AA$ (6 pixels).}\label{fig:sigmanscale}
\end{figure}

\subsubsection{Noise properties}\label{sec:noisemuse}

During data reduction, the detector noise is propagated through the different reduction steps and finally combined into a cube that contains the pixel variance. As a result of the several transformations undergone by the detector pixels, including interpolation on a final data cube, the resulting pipeline variance  does not accurately reproduce the effective standard deviation of the voxel (volumetric pixels) inside the final data cube, but it still reflects the relative variation of the noise as a function of position and wavelength. This is shown explicitly in Fig.~\ref{fig:rmshisto}, where we show the flux distribution of voxels ($f_{\rm vox}$) in the range $4900-5500~\rm \AA$ (i.e. the range that covers Ly$\alpha$ at the redshifts of interest) normalised by the pipeline error in each voxel ($\sigma_1$). Once sources are masked, the distribution is expected to approximate a Gaussian with standard deviation of unity.  Fig.~\ref{fig:rmshisto} shows instead that the distribution of $f_{\rm vox}/\sigma_1$ (red line) from all the MUSE data cubes combined has a characteristic width of $\approx 1.24$, implying that the pipeline error underestimates the true flux standard deviation.

One way to mitigate this effect is to renormalise the pipeline pixel variance by a wavelength-dependent factor computed by comparing the pipeline variance with the flux distribution in each layer of a data cube. This technique indeed yields a rescaled variance that more closely reproduces the pixel noise, as shown in Fig.~\ref{fig:rmshisto} (green line). 
However, a similar scaling would not be appropriate for data that are combined with a median rather than a mean, as the error on the median is known to be $\approx 1.25$ times that of the mean. We therefore proceed by bootstrapping pixels in individual exposures using the resampled cubes after the {\sc cubex} post-processing to reconstruct an estimate of the noise for the final mean, median, and half-exposure cubes. For this, we use 20,000 realisations which we find to be enough for convergence (we explicitly test for convergence by recomputing the noise using 10,000 and 100,000 samples). 

As the quality of the reconstruction is ultimately limited by the small number of individual exposures, we then use the bootstrap estimates to derive a wavelength-dependant scaling coefficient that we then apply to the pipeline variance. In this way, we find a better estimate of the amplitude of the pixel variance, while retaining the relative variation as a function of wavelength and position of the pipeline variance. Fig.~\ref{fig:rmshisto} shows that indeed, once the variance has been rescaled following this procedure,
the $f_{\rm pix}/\sigma_1$ distributions are characterised by a standard deviation close to unity, showing that we are able to correctly describe (to within a few percent) the noise properties at the pixel level.     

While the above procedure yields a better estimate of the pixel rms, the resampling of individual pixels inside a final datacube further introduces correlations that are likely to result in a weaker scaling of the signal-to-noise as a function of number of pixels $N$ compared to the theoretical estimate from the propagation of the pixel error, $\sigma_{\rm N} = \sigma_1 \sqrt N$. This effect, commonly seen in drizzled images \citep[e.g.][]{fumagalli2014}, would therefore result in an underestimate of the effective noise inside an aperture ($\sigma_{\rm eff}$), and hence would produce an overestimate of the real $S/N$ of a source.

To model the change in the noise as a function of number of pixels in a detection aperture, we compute the effective noise $\sigma_{\rm eff}$ as the standard deviation of fluxes from empty regions (i.e. excluding sources) across the MUSE cubes assuming cubic apertures of $5~\rm \AA$ (4 spectral pixels) and a variable aperture size in the spatial direction. Fig.~\ref{fig:sigmanscale} shows a comparison between the effective noise, $\sigma_{\rm eff}$, as a function of aperture size with respect to the expectation from propagating the formal error, $\sigma_{\rm N}$.
As is evident, the effective noise is already $\approx 50\%$ higher than the formal error for an aperture of $\approx 1~\rm arcsec$ on a side. A quadratic fit to the median ratios across the different cubes yields an expression of the form  
$\sigma_{\rm eff}/\sigma_{\rm N} = 1.429 + 0.091 L - 0.008 L^2$, with $L$ the diameter of the aperture in arcsec. This fitting function is useful to estimate the corrected $S/N$ of detections, as done in Sect.~\ref{sec:galaxies}.
As shown in Fig.~\ref{fig:sigmanscale}, although this correction has an additional dependence on the width of the spectral window, it is accurate to $\approx 5\%$ for typical widths of narrow emission lines (few \AA) in galaxies. 

\begin{figure*}
    \centering
    \includegraphics[width=\textwidth, trim = 0cm 0.5cm 0cm 0cm, clip]{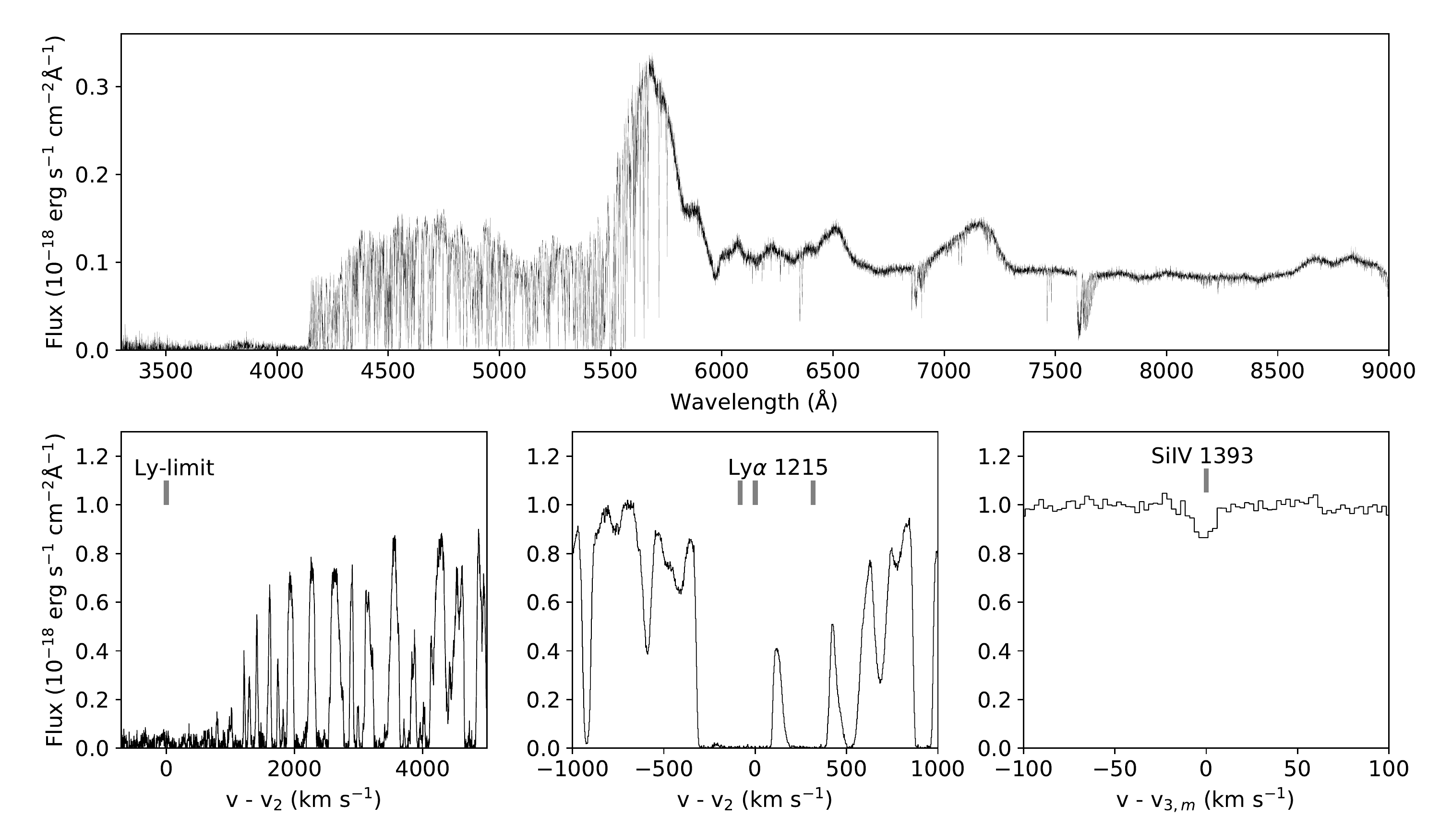}
    \caption{Spectrum of the quasar \qname. \textit{Top:} Full X-SHOOTER spectrum. \textit{Bottom left:} Normalised UVES spectrum showing the Lyman limit at 4150~\AA\ and a portion of the Lyman series. \textit{Middle:} Ly$\alpha$ absorption line. The grey vertical line marks the position of the Ly$\alpha$ absorption line arising from the central absorption component. The x-axis in this and the left panel show the velocity relative to the redshift of the central component of the LLS, $z_2 = 3.525285 \pm 0.000012$. 
      \textit{Right:} The weak \ion{Si}{IV} absorption associated with the LLS. The x-axis shows the velocity relative to the redshift of the associated metals as measured by \citet{crighton2016}, $z_{3,m}=3.530217\pm0.000002$. }
    \label{fig:quasar}
\end{figure*}

\begin{figure*}
    \centering
    \includegraphics[width=\textwidth, trim = 0cm 0.5cm 0cm 0cm, clip]{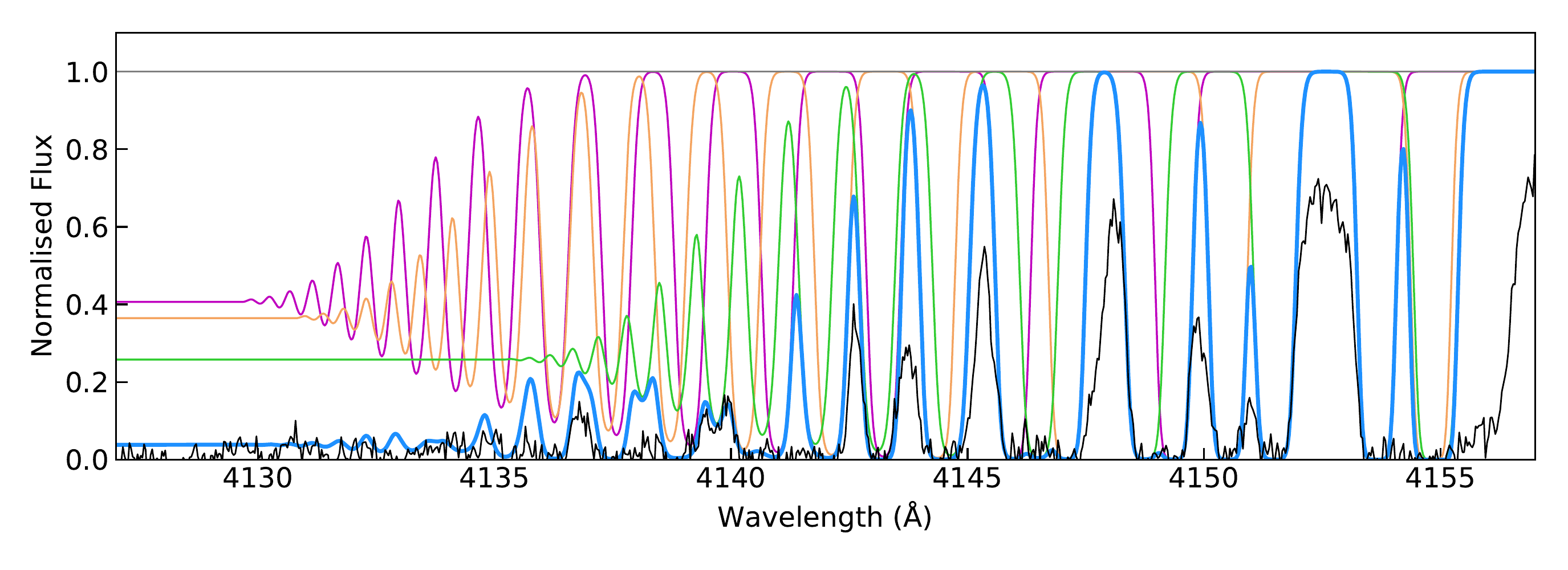}
    \caption[width=\textwidth]{Normalised UVES spectrum for the quasar \qname~ at the Lyman limit of LLS1249 (black) as in \citet{crighton2016}. Fits for the three absorption components of the LLS are shown in magenta, orange and green using the column densities, redshifts and Doppler parameter \textit{b} from their paper. The solid blue line shows the combined spectrum resulting from these three components.}
    \label{fig:lls}
\end{figure*}

\begin{figure}
    \centering
    \includegraphics[width=0.5\textwidth, trim = 0.5cm 0.5cm 0cm 0cm, clip]{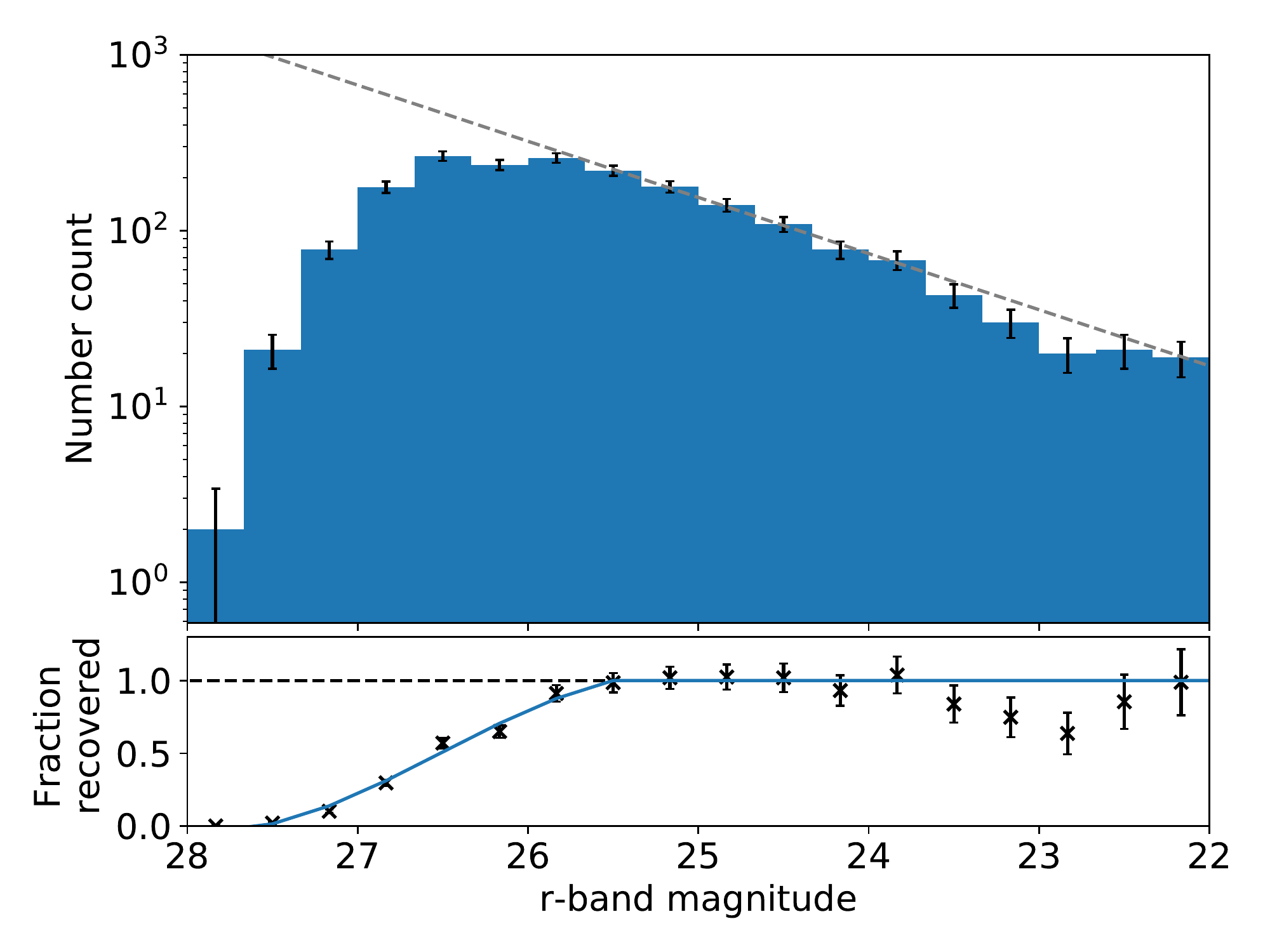}
    \caption{{\it (Top)}: Number of galaxies extracted as a function of $r$-band magnitude from all the sightlines which have 5-hour MUSE exposures. The grey dashed line shows the expected number of galaxies determined from a linear fit to the region between $m_r>23$~mag and $m_r<25.5$~mag. {\it (Bottom)}: Fraction of galaxies recovered relative to the expected number as a function of $r$-band magnitude. Our survey achieves 100\% completeness down to 25.5 mag and 90\% completeness at 26.3 mag.
}
    \label{fig:numcounts}
\end{figure}

\section{Strong absorption line systems}\label{sec:als}

\subsection{Methodology}

All the quasars targeted with our MUSE observations have high-resolution archival spectra, often complemented by moderate resolution spectroscopy, from which we identify and characterise absorption line systems along the quasar line of sight. To ensure that our selection is unbiased with respect to metal content and kinematics, we identify the systems by searching the spectra for features indicative of \ion{H}{I} absorption, including damped Ly$\alpha$ absorption lines and the characteristic break in the quasar continuum at the Lyman limit of optically-thick absorbers. The presence of a LLS is then confirmed by overlaying Voigt profiles for the Lyman series and, although not a requirement for selection, by the presence of metal transitions (e.g. CII, CIV, SiII, SiII, OVI). By visually adjusting these fits until we find the best match to the observed spectrum, we determine the redshift of the absorption line system. In order to confirm the identification of the systems, the analysis for each of the sightlines was repeated independently by two authors (EKL, MF). 
Following this procedure, we identified \nlls\ absorbers along the \nqso\ sightlines, as detailed in Table~\ref{tab:sample}.

Having identified the absorbers, we characterise their properties using an interactive fitting procedure from the {\sc pyigm} python package\footnote{\url{https://github.com/pyigm/pyigm}.}. For systems with high \ion{H}{I} column densities, we use the \textit{fitdla} routine to model the damped wings of the Ly$\alpha$ lines. In these cases, we determine the \ion{H}{I} column density by firstly visually fitting the continuum level and then adjusting the column density and Doppler parameter, $b$, to find the best fit value. 
For systems with lower column densities, the Ly$\alpha$ line is not damped, and therefore the line profile cannot be used alone to determine the column density. Following common procedures \citep[e.g.][]{prochaska2015}, we instead use the break in the emission at the Lyman Limit (912~\AA\ in the rest-frame) and the full Lyman series to obtain the best constraint on the \ion{H}{I} column density. Also in this case, we first fit the continuum level at wavelengths above the break and then adjust the column density to find the best fitting profile. Details on the column densities of the entire sample of absorbers, together with a characterisation of their metal content, will be presented in forthcoming publications together with the analysis of the MUSE data. 

\subsection{Application to \qname}
The MAGG survey targets a large sample of LLSs which are believed to be a signpost of the CGM at z$\sim$2-3 where active galaxy formation is taking place \citep[e.g.][]{sargent1989,fumagalli2011}.
In this paper, we focus on the first sightline studied from MAGG, chosen due to the presence of a very metal-poor LLS along the line of sight, as an example of the results that this survey will produce. To date, only a very limited number of near-pristine and pristine LLSs are known \citep[e.g.][]{fumagalli2011sci,crighton2016,cooke2017,robert2019}, and their environment has been studied only in two instances \citep{fumagalli2016}.
Despite their small number, these systems are of considerable interest
being possible candidates for clouds that may have been enriched solely by the very first generation of stars \citep[e.g.][]{welsh2019}. These clouds could also be the antecedents of the lowest mass galaxies in the local Universe, and are compelling candidates of the elusive cold-mode accretion.

The field we study is centred on the quasar \qname\ which is at a redshift of $z\approx  3.634$ \citep[see also Table~\ref{tab:sample}]{schneider2003} and which intersects a previously identified LLS (hereafter LLS1249) at $z \approx 3.53$ \citep{crighton2016}, see Figure~\ref{fig:quasar}. High resolution archival spectroscopy exists for this quasar from UVES (programme 075.A-0464, P.I. Kim), HIRES (U157Hb, P.I. Prochaska) and X-SHOOTER, as detailed in Table~\ref{tab:qsospectra}. From these spectra, \citet{crighton2016} determined that LLS1249 is composed of three \ion{H}{I} absorption features spanning  $\approx 400~\rm km~s^{-1}$. Fits for these three absorption components to the normalised UVES spectrum are shown in Figure~\ref{fig:lls} and they agree well with the observational data around the Lyman limit. 
\citet{crighton2016} determined redshifts and column densities for these components as follows. The central component of the LLS (component 2) is at $z\approx 3.5252$ and has a hydrogen column density of $\log (N_{\rm HI}/\rm cm^{-2})=17.20 \pm 0.03$. We use this component as the reference for comparison in our work. Component 1 lies at z = 3.5240 ($\Delta v = -83~\rm km~s^{-1}$) with  $\log (N_{\rm HI}/\rm cm^{-2})=17.15 \pm 0.04$ while the third component is at z = 3.5301 ($\Delta v = +317 ~\rm km~s^{-1}$) and has a hydrogen column density of $\log (N_{\rm HI}/\rm cm^{-2})=17.33 \pm 0.03$. Together, these components result in a LLS with a combined column density of $\log (N_{\rm HI}/\rm cm^{-2})=17.74 \pm 0.03$. 

In this sightline, \citet{crighton2016} detected doubly and triply ionised carbon in all components with low column densities ($\approx 10^{12.5}~\rm cm^{-2}$). Silicon is also detected in the third component (Figure~\ref{fig:quasar}). While two of the components did not have a sufficient number of identified metal transitions to reliably determine the ionisation parameters, and hence their metallicity, it was possible to infer an abundance for the third component. Through photoionisation modelling, \citet{crighton2016} found a metallicity of $\rm [Si/H] = -3.41 \pm 0.26$, 
making this system one of the most metal-poor LLSs known at these redshifts.
While the exact value is sensitive to the photoionisation modelling, the lack of detection in strong metal transitions robustly places this system below a metallicity of $\approx 1/1000$ solar.   
Given this extremely low metallicity, these authors concluded that LLS1249 may have been enriched by only the very first stars in the Universe, making this absorber a candidate remnant of Population III stars. This is supported by the observed C/Si ratio which, while being consistent with later Population II enrichment, is a distinctive signature of models of gas enrichment from Population III stars.

\section{Detection of galaxies in MUSE data}\label{sec:galaxies}

To achieve the main goal of our survey, for each of the quasar fields in our sample, we need to identify galaxies associated with the DLAs and LLSs by conducting searches for emission line objects as well as continuum sources in the deep white-light images reconstructed from the MUSE data cubes.
In the following, we present the procedure we adopt in MAGG, which we apply in this paper to the field \qname.

\begin{figure*}
    \includegraphics[width=0.7\textwidth, trim = 0cm 0.5cm 0cm 0cm, clip]{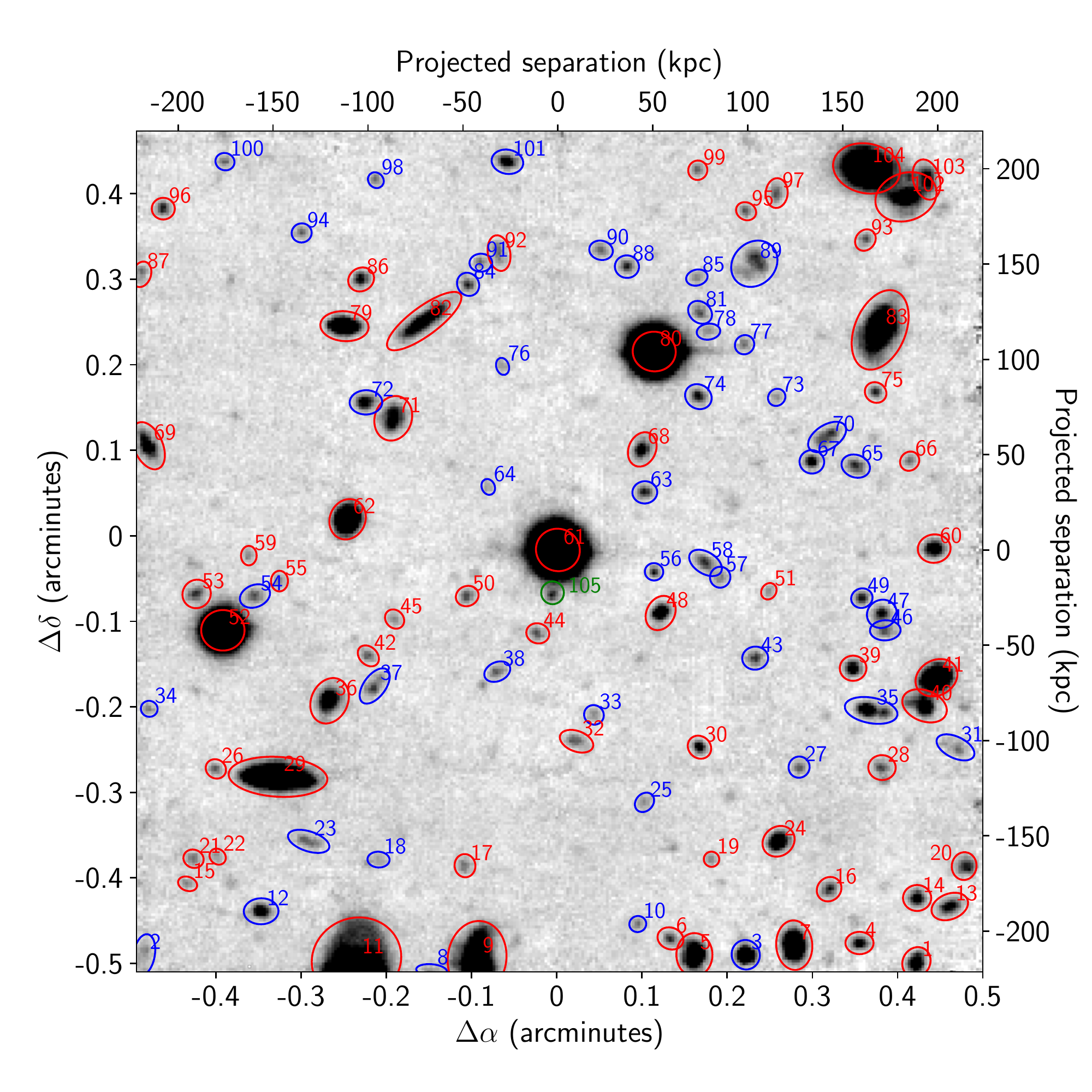}
    \caption{White-light image of the MUSE FOV, centred on the position of the quasar \qname. Continuum-detected sources with reliable spectroscopic redshifts from {\sc MARZ} are highlighted with a red aperture and a source ID. Sources with no confident redshifts are indicated in blue. Remaining features without an aperture fell below the $\rm S/N >2$ constraint and were not extracted. The source shown in green south of the quasar was not detected by {\sc SExtractor} due to blending but was instead extracted manually. 
    }
    \label{fig:continuum_sources}
\end{figure*}

\begin{figure}
    \centering
    \includegraphics[width=0.5\textwidth, trim = 0cm 0.5cm 0cm 0cm, clip]{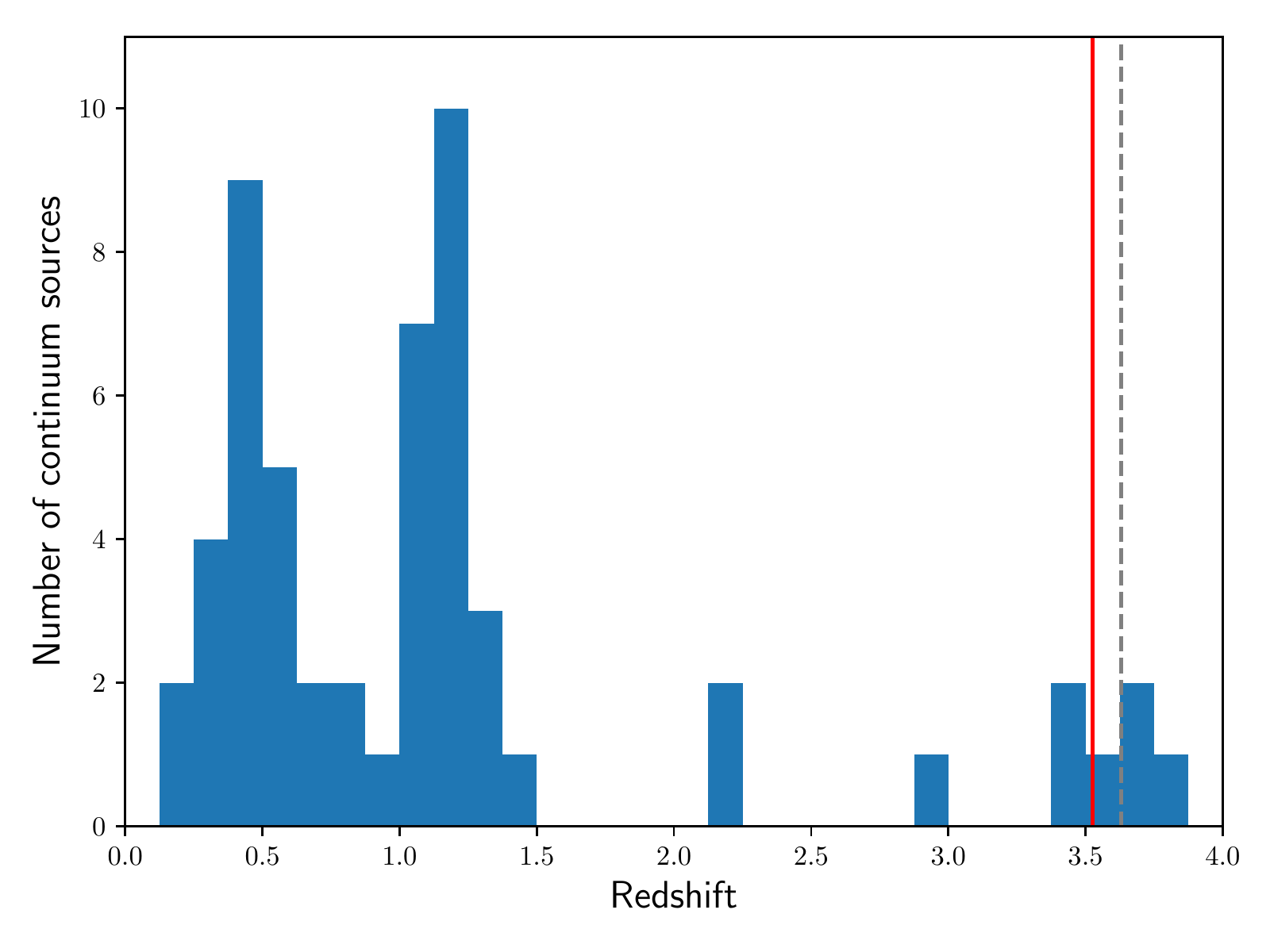}
    \caption{Redshift distribution of continuum-detected sources with a confidence flag of 2 or more in the MUSE cube of the \qname\ quasar. The grey dashed line shows the redshifts of the quasar ($z\approx 3.63$) and the red one shows the one of the LLS observed in the quasar spectrum at $z \approx 3.525$.}
    \label{fig:z_dist}
\end{figure}

\subsection{Continuum-detected sources}\label{subsec:continuum}

\subsubsection{Methodology}

To identify continuum sources, we extract objects using {\sc SExtractor} \citep{bertin1996} on the white light images reconstructed from the MUSE datacubes. For source extraction, we use a minimum area of 6 pixels each, a detection threshold of twice the rms background level and a minimum deblending parameter {\sc DEBLEND\_CONT} of 0.0001 to best detect sources in proximity of bright objects, such as the quasar or stars within the FOV. To assess the completeness of our source extraction procedure, we derive number counts of continuum-detected sources as a function of $r$-band magnitude, shown in Figure~\ref{fig:numcounts}, for all the sightlines which have 5-hour MUSE observations. The handful of archival sightlines with longer exposures will yield correspondingly deeper completeness limits.

Assuming an underlying power-law distribution, we fit the linear region between $m_r>23.5$~mag and $m_r<25.5$~mag, which we extrapolate to determine the expected number of galaxies at fainter magnitudes \citep[see e.g.][]{fumagalli2014}. We use this estimate to then calculate the fraction recovered and determine that our sample is $\approx 100\%$ complete down to an $r$-band magnitude of $\approx 25.5$~mag and $\approx 90\%$ complete down to $\approx 26.3$~mag.

Following the identification of continuum sources, we use the segmentation map to define Kron apertures (or circular apertures, in case of compact sources) on the datacubes from which we extract spectra for the selected objects. We use the {\sc MARZ} tool \citep{hinton2016} to classify the source redshifts from the extracted spectra. Sources below this magnitude limit, where 
we are only 55\% complete, have spectra of insufficient quality to attempt a redshift using cross-correlations (see below). However, any galaxy with continuum below this magnitude limit showing bright emission lines will still be included in our catalogue following the search for emission lines described in Sect.~\ref{subsec:emission}.

In MAGG, we use the M. Fossati fork\footnote{\url{https://matteofox.github.io/Marz/}.} of {\sc MARZ}, which includes additional high-redshift templates and high-resolution templates well-suited for MUSE data. 
With {\sc MARZ}, the 1D spectra for each source are compared to galaxy and stellar templates with a cross-correlation, and the results are visually inspected to confirm emission and absorption lines and characteristic broad continuum features. For each redshift measurement, we assign a confidence flag, ranging from 1 to 4 with stellar sources graded 6. In assigning these grades, we follow the categorisation used in \citet{bielby2019}, namely: 
\begin{itemize}
    \item[1.]  Low S/N spectrum with no clearly identifiable features which can yield a confident redshift measurement;  
    \item[2.]  Single emission line with low S/N continuum and weak or no other identifiable lines, for which the redshift is uncertain; 
    \item[3.] One strong emission or absorption line with some additional low S/N emission or absorption features for which we can determine a confident redshift;
    \item[4.] Multiple high S/N emission or absorption lines, which yield an accurate redshift.
\end{itemize}

\subsubsection{Application to \qname}

Using the method described above, we extract 104 continuum-detected sources in the 1 arcmin$^2$ FOV around the  \qname\ quasar. Figure~\ref{fig:continuum_sources} shows a white-light image obtained by collapsing the mean datacube centred on the quasar, with apertures highlighting the sources extracted using {\sc{sextractor}}. Continuum-detected sources  are highlighted with an aperture and source ID corresponding to that in Table~\ref{tab:continuum_sources}. Sources are identified by a red aperture if they have reliable redshifts and blue if they do not. The redshifts are measured using {\sc Marz} as described above. In the end, we derive 55 redshifts with flag $\ge 2$, with an 80\% completeness for sources with $m_{r} < 25$~mag. 

Properties of continuum-detected sources, including redshifts, are listed in Table~\ref{tab:continuum_sources}, while Figure~\ref{fig:z_dist} presents the redshift distribution of the sources identified within the MUSE datacube for \qname. Continuum sources which were flagged as 1 in {\sc MARZ} are excluded as their spectra were either low S/N or they showed no clear spectral features and hence their redshifts are undetermined. Our redshift analysis reveals several emission line galaxies at $z<1.5$ and at $z>3$, with only a handful of sources lying in the ``MUSE redshift desert'' where [OII] and Ly$\alpha$ are not in the accessible wavelength range but redshifts can still be established based on absorption lines. No continuum sources are found within $1000~\rm km~s^{-1}$ ($\Delta z \simeq 0.003$) of the central \ion{H}{I} component of LLS1249.

\begin{table*}
\centering
\caption{Continuum sources in the MUSE FOV extracted by {\sc{Sextractor}} with S/N$>2$ and m$_{r} <27$ mag. Column 1 shows the source ID; column 2 shows the source name; columns 3 and 4 list the right ascension and declination of the sources followed by the r-band magnitude of the source in column 5 with its associated error (column 6). The redshifts obtained using {\sc{MARZ}} are shown in column 7 followed by their confidence (column 8). A confidence flag of 4 indicates our high confidence sources while flag 1 is for the lowest confidence and the associated redshifts are unreliable.  The full table is included as online only material.}\label{tab:continuum_sources}
\begin{tabular}{cccccccc}
\hline
ID & Name  & R.A. & Dec. & $m_{\rm r}$ & $\sigma_{m_{\rm r}}$ & Redshift & Confidence \\
\hline
\hline
1 & MUSEJ124955.56-015957.6 & 192.48148 & -1.999347 & 24.4 & 0.1 & 1.3644 & 3 \\
2 & MUSEJ124959.18-015957.1 & 192.49659 & -1.999203 & 26.0 & 0.3 & - & 1 \\
3 & MUSEJ124956.36-015957.1 & 192.48482 & -1.999202 & 23.6 & 0.1 & - & 1 \\
4 & MUSEJ124955.82-015956.3 & 192.48259 & -1.998974 & 25.4 & 0.2 & 1.36405 & 3 \\
5 & MUSEJ124956.60-015957.1 & 192.48582 & -1.999203 & 23.4 & 0.09 & 1.10872 & 4 \\

\hline
\end{tabular}
\end{table*}

Figure~\ref{fig:continuum_sources} further shows a source south of the quasar (green aperture) which was not extracted by {\sc{sextractor}} due to heavy blending. In this case, we manually extract a spectrum of this source by drawing a circular aperture around its position. Due to its close proximity to the quasar, the extracted spectrum contains emission from the tail of the quasar point spread function (PSF). To subtract this residual quasar emission, we draw an annulus centred on the quasar at the same radius as the source aperture, but excluding the region where the source lies. In each wavelength slice, we subtracted the median flux from this annulus from each pixel in the source region. The resulting spectrum is then analysed in {\sc MARZ} as for all the other detected continuum sources. However, the spectrum is found to be at too low S/N with no prominent features. Therefore, we are not able to determine a reliable redshift for this object, to which  we assign a confidence flag of 1. 

\begin{figure}
    \centering
    \includegraphics[width=0.5\textwidth]{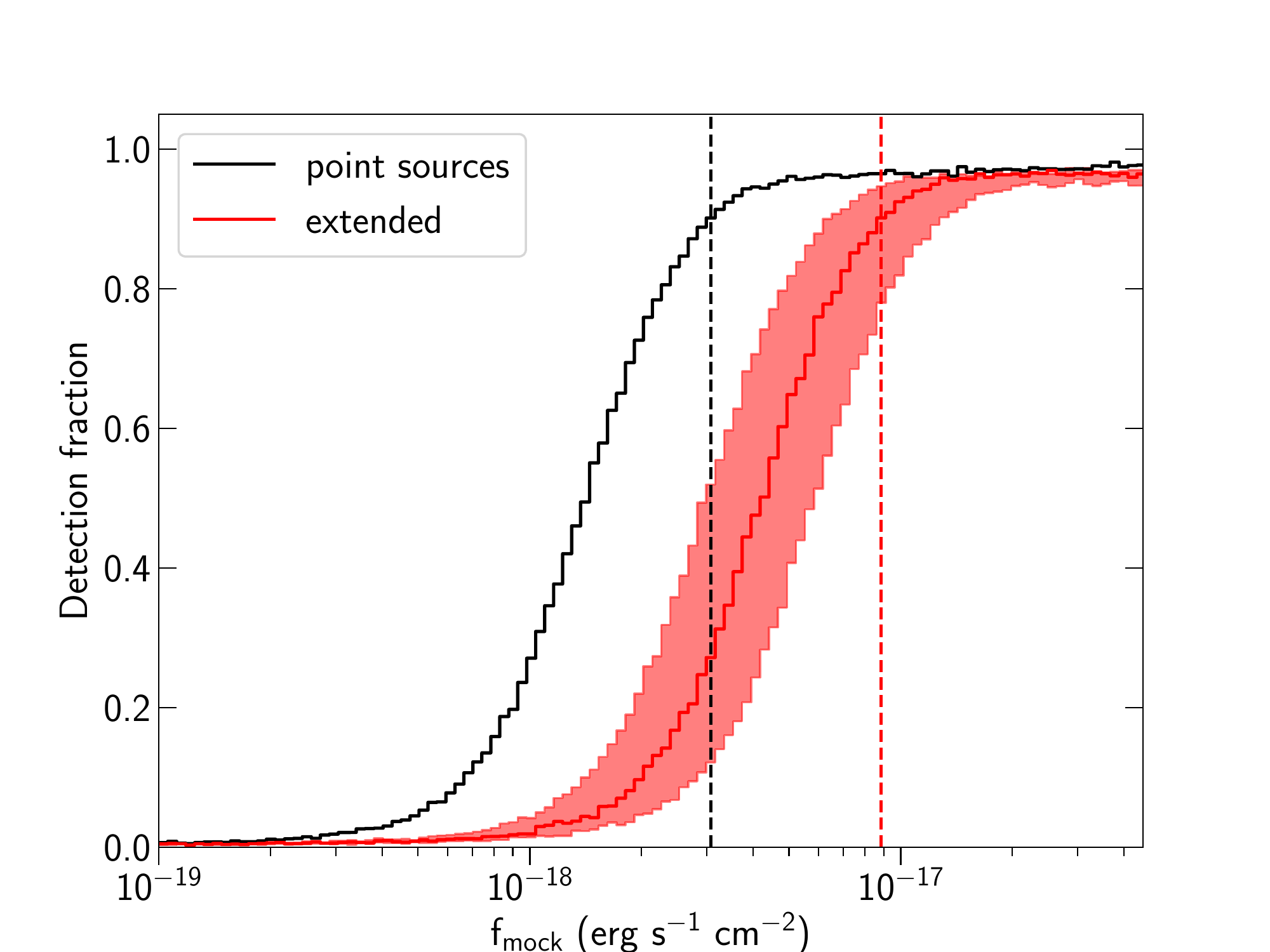}
    \caption{Fraction of simulated line emitters detected as a function of flux. For point sources (black), we are 90\% complete down to $\approx 3\times 10^{-18}~\rm erg~s^{-1}~cm^{-2}$ (vertical line). The solid red line shows the recovered fraction for exponential disks with scale lengths of 3.5 kpc. For these sources, we are 90\% complete to $\approx 9\times 10^{-18}~\rm erg~s^{-1}~cm^{-2}$. The red shaded region indicates the completeness for extended sources with scale lengths ranging from 2 kpc to 5 kpc.}
    \label{fig:LAEcompleteness}
\end{figure}

\subsection{Emission line sources}\label{subsec:emission}

\subsubsection{Methodology}

In addition to the redshift survey of continuum-detected galaxies, we conduct a search for emission line galaxies, and in particular Ly$\alpha$ emitters (LAEs) within the MUSE FOV. This extends our ability to find associations with the DLAs and LLSs observed in the quasar spectra by including sources that are faint in the continuum but sufficiently bright in their emission lines.

Our first step is to cut down (for computational efficiency) the mean cube obtained using the {\sc CubEx}  reduction (see Section~\ref{sec:cubexredux}) to the wavelength of Ly$\alpha$ at the redshift of the absorbers plus 300 wavelength channels (375\AA) either side as in \citet{mackenzie2019}. We create similarly trimmed median and data cubes containing half of the exposures for quality control. We then use {\sc CubEx} to subtract the PSF of the quasar and remove continuum sources from this reduced cube using the procedure described in \citet{arrigoni2019}.
At this stage, data at velocities overlapping with the location of DLAs and LLSs are masked in order to prevent emitters being subtracted close to the absorption system redshift.The continuum  within this velocity range is calculated using an extrapolation of the continuum in the unmasked wavelengths and subtracted.

{\sc CubEx} is then run on the continuum-subtracted cube to search for and extract potential line emission
galaxies. The spatial positions of continuum-detected sources that have known redshifts (confidence $\geq$~2 from the {\sc MARZ} analysis) are masked.  This extraction involves convolving the detection cubes with a two-pixel boxcar in the spatial direction and defining groups from any set of connected voxels that all have an individual $\rm S/N > 3$.
In order to be selected as a candidate line emitter, the group must then satisfy the following criteria: (i) the group consists of more than 27 voxels, (ii) in at least one spatial position within the detection, the pixels span at least 3 wavelength channels ($>3.75\AA$), (iii) in order to exclude residuals from continuum sources, the group must not span more than 20 wavelength channels.

Residual cosmic rays that are not fully removed by the $\sigma$-clipping algorithm adopted when combining the data are one of the most common contaminants in our extraction. Although cosmic rays are typically narrow in both the spatial and spectral directions, they can occasionally meet the above criteria and are flagged as detections. To remove them from the source catalogues, we add conditions on the $S/N$ in the independent coadds, as cosmic rays are present in only one exposure and thus appear detected in one half of the data but not the other half. In this way, we are able to effectively remove cosmic rays from the catalogue.

 The candidate line emitters are then classified into two confidence groups  \citep[see also][]{mackenzie2019} based on the integrated $S/N$ ($ISN$) of the entire source, corrected for the effective noise as described in Section~\ref{sec:noisemuse}. The first class contains sources with an $ISN>7$ and consists of our highest purity sample, at the expense of a lower completeness. The second class includes sources with an $ISN>5$, 
 extending the completeness at the expense of the purity. Indeed, classification is not always unambiguous for sources that approach the detection limit.  This class thus contains, especially around $ISN\approx 5$, candidate emitters for which deeper data are needed to confirm their nature.
 For both of these classes, we require that the difference in $S/N$ between each of the independent coadds is less than 50\%, which is found to be a good discriminant between sources and residual artefacts. Finally, for every source, we also monitor the $ISN$ in the median coadd (which contains similar information to the one captured by the independent coadd).
 
Following the extraction and classification of line emitters, the 3D segmentation maps for each of the sources are projected into the spatial  dimension and used to extract a spectrum across the full MUSE wavelength range ($4650\AA - 9300\AA$ in extended mode). As a further quality control, we visually inspect the sources using spectra and images extracted from each of the cubes (mean, median, independent coadds and detection datacubes). The 3D segmentation maps are also checked to ensure that the identified detections appear to be either a point source or an extended source without any extreme elongation particularly in wavelength which is unlikely to be physical. At this stage, we also inspect the spectra for each source to separate LAEs from other lines such as [OII] (partially resolved in MUSE) or continuum source residuals. [OII] emitters are catalogued separately for use in future papers. 

To test the completeness of our search for line emitters, we repeat the extraction process described above 1000 times using synthetic sources drawn from a uniform distribution of fluxes between $10^{-19}$ and $5\times10^{-17}~\rm erg~s^{-1}~cm^{-2}$. In each run, we inject 500 sources (to avoid source confusion) into the mean datacube at random spatial positions and a random wavelength between 5000~\AA\ and 5350~\AA,
which is the range of interest for our search and is free from bright sky lines. 
At first, we inject point-like sources that are defined by a 2D Gaussian in the spatial dimensions with a FWHM of 0.7~arcsec and a 1D Gaussian in the spectral dimension with a line spread function of FWHM of 2.5\AA. These synthetic cubes are then processed in the same way as the real data,  and we use the final catalogues to determine how many of the mock sources are recovered. 
Figure~\ref{fig:LAEcompleteness} shows this fraction of recovered sources as a function of flux. For point-like sources (black), we recover 90\% of our injected sources down to $\approx 3\times 10^{-18}~\rm erg~s^{-1}~cm^{-2}$.

As discussed in the literature \citep[e.g.][]{herenz2019}, this test is somewhat optimistic, as extended sources are more difficult to recover due to their extended surface brightness profile. We therefore repeat this exercise using extended sources in place of the point sources which are more comparable to the observed LAEs. The injected sources are modelled using exponential disks in the spatial directions with intrinsic face-on scale lengths of 2, 3.5 and 5 kpc and a 1D Gaussian in the spectral direction with a FWHM of 2.5\AA, as for the point-like sources. These models are spatially convolved with the seeing.  Using extended sources reduces our completeness at any given flux, with a 90\% completeness at $\approx 9\times 10^{-18}~\rm erg~s^{-1}~cm~^{-2}$ for extended sources with a scale length of 3.5 kpc.
For a fixed flux of $5\times 10^{-18}~\rm erg~s^{-1}~cm~^{-2}$, we recover 96\% of point-like sources and 65\% of 3.5 kpc extended sources.

\begin{figure*}
    \centering
    \includegraphics[scale=1.2, trim= 1cm 0cm 1cm 0cm, clip]{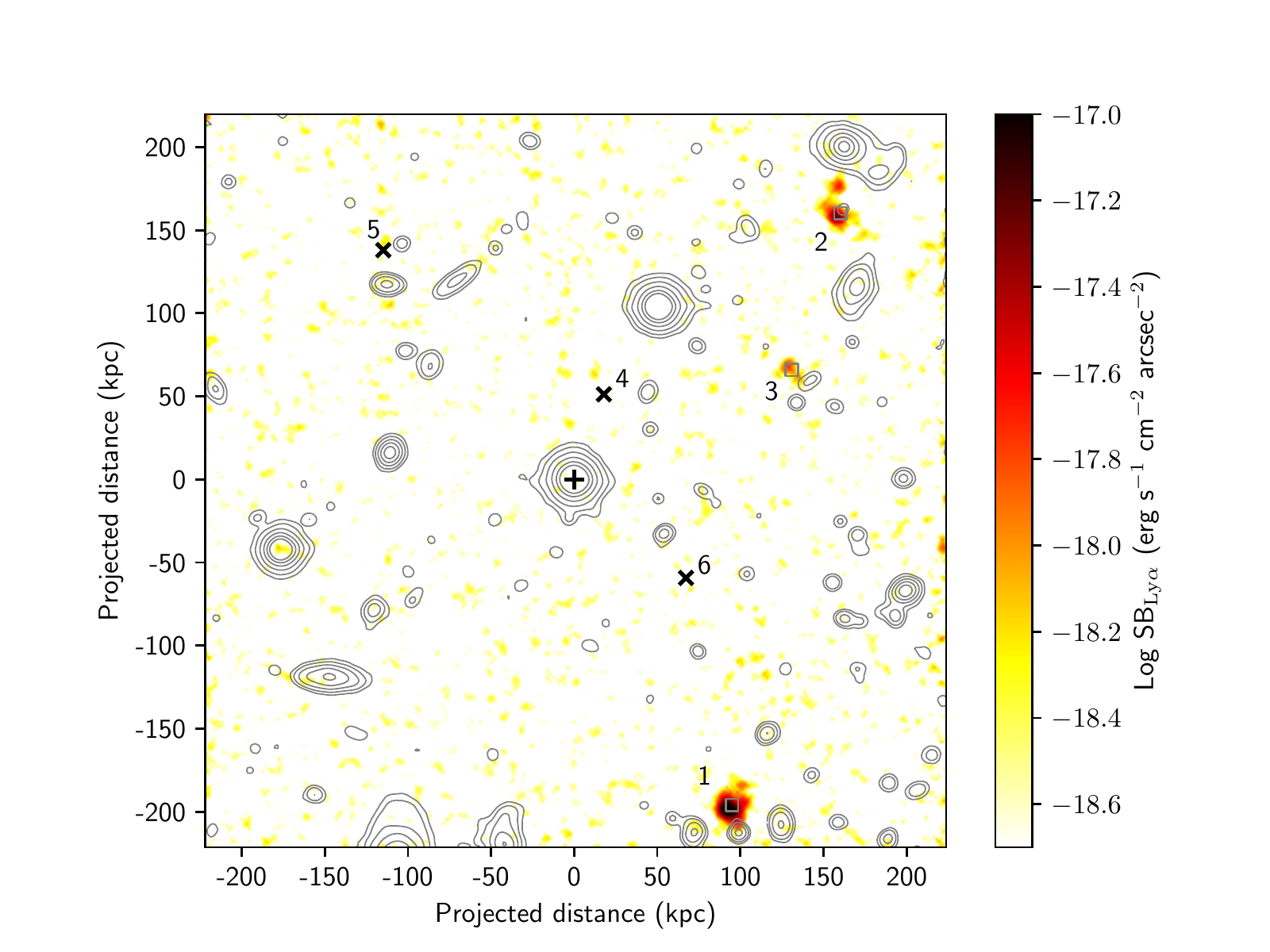}
    \caption{A narrow-band image of Ly$\alpha$ emission at the redshift of LLS1249 reconstructed from the MUSE data cube by optimally integrating flux across emission lines. We detect three LAEs in the MUSE FOV within $200~\rm km~s^{-1}$ of the LLS redshift, as marked by the grey squares.  The black plus sign in the centre of the image marks the position of the quasar while the black crosses mark the positions of the lower confidence emitters described in Table~\ref{tab:emitters}. We overlay black contours to show the continuum sources (unrelated to LLS1249), as seen in Figure~\ref{fig:continuum_sources}, at levels of 22, 23, 24, 25, 26 and 27 mag arcsec$^{-2}$. The Ly$\alpha$ emission has been smoothed using a top-hat kernel of width 0.4~arcsec.}
    \label{fig:LAE_map}
\end{figure*}

\begin{figure*}
    \centering
    \includegraphics[width=\textwidth, trim= 0cm 0.5cm 0cm 0cm, clip]{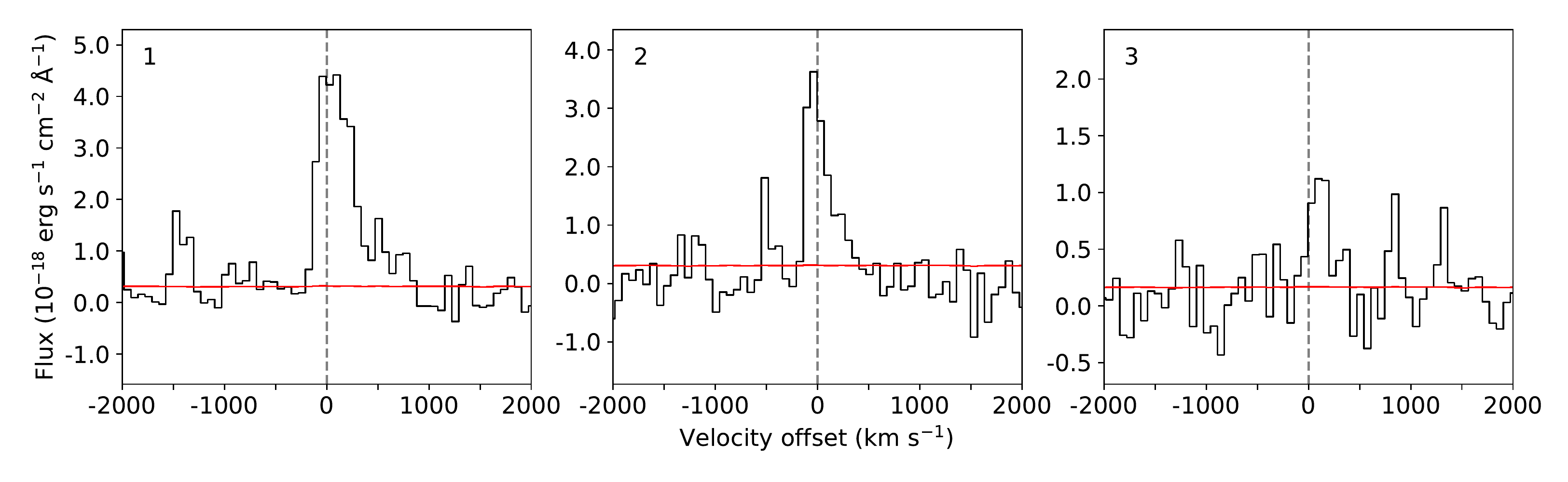}
    \caption{The Ly$\alpha$ emission extracted from the MUSE datacubes for each of the three high confidence sources detected around the LLS in \qname. These spectra extracted from the spatial regions defined by their segmentation maps. The velocity is shown relative to the LLS at z=3.52522 and the error is shown by the solid red line.}
    \label{fig:LAE_spectra}
\end{figure*}

\subsubsection{Application to \qname}

Applying the method described above for the detection of emission line sources to the field of \qname, we extracted potential line emitters from the mean datacube using {\sc CubEx} in a window of $\pm 1000~\rm km~s^{-1}$ centred on the central component of the $z\approx 3.53$ LLS. This window is sufficiently large to encompass galaxies in the near environment of the absorption line systems offset both due to peculiar velocity and the Hubble flow.

Following the search, we classify sources into the confidence groups as described above. We then visually inspect the cut-out images from the mean, median and independent cubes with half-exposures as well as 1D spectra extracted from the spatial regions defined by their segmentation maps and their shape in these segmentation maps to rule out any sources with unusual shapes. 
Sources which do not appear to be real based on these checks are excluded from our catalogue. 
In the end, we identify 3 detections in group 1 and 3 possible detections in group 2.
For this sightline, we find no foreground or background sources (e.g. [OII] emitters) within the velocity window of LLS1249. Our final catalogue of sources associated with LLS1249 is shown in Table~\ref{tab:emitters}.

\begin{table*}
\centering
\caption{Line emitters extracted within 1000 km s$^{-1}$ of the LLS at $z\approx 3.525$. The table lists: the ID; the right ascension; the declination; the line flux and luminosity with associated errors; the integrated $S/N$ of the source; the confidence class based on the $S/N$ where class 1 are our highest confidence sources; the redshift; the velocity offset relative to the central component of the LLS.  }\label{tab:emitters}
\begin{tabular}{ccccccccc}
\hline
ID & R.A. & Dec. & F$_{\rm line}$  & L$_{\rm line}$ & ISN & Class & Redshift & Velocity offset \\
& (hrs) & (deg) &  (10$^{-18}$ erg s$^{-1}$ cm$^{-2}$) & (10$^{41}$ erg s$^{-1}$) &&  & & (km s$^{-1}$)\\ 
\hline
\hline
1 & 12:49:56.409 &	-01:59:55.19 &	26.1 $\pm~1.1$ & 31.2 $\pm~1.3$ &	23.7 &	1 &   3.52727 & 136 \\
2 & 12:49:55.825 &	-01:59:07.53 &	11.3 $\pm~0.8$ & 13.5 $\pm~1.0$ &	14.6 &	1 &   3.52558 & 24 \\
3 & 12:49:56.091 &	-01:59:20.13 &	2.6 $\pm$~0.3 & 3.1 $\pm$~0.3 	 &	7.39 &	1 &   3.52670 & 98 \\
4 & 12:49:57.096 &	-01:59:22.08 &	1.5 $\pm~0.3$ & 1.8 $\pm~0.3$ 	 &	5.11 &	2 &   3.51893 & -417 \\ 
5 & 12:49:58.276 &	-01:59:10.42 &	1.4 $\pm~0.3$ & 1.7 $\pm~0.3$ 	 &	5.14 &	2 &   3.53316 & 526 \\
6 & 12:49:56.650 &	-01:59:36.85 &	1.5 $\pm~0.3$ & 1.8 $\pm~0.3$ 	 &	5.96 &	2 &   3.54012 & 987 \\

\hline
\end{tabular}
\end{table*}

The three newly-identified high confidence Ly$\alpha$ emitters found in close proximity to the redshift of LLS1249 are shown in Figure~\ref{fig:LAE_map}. In this figure, we overlay contours showing the position of all the continuum sources detected in the field. None of these sources are associated with the 3 LAEs, including the source spatially coincident with LAE 2, which is a lower redshift galaxy. We also show the location of the further 3 candidate detections in class 2 as black crosses. Figure~\ref{fig:LAE_spectra} shows the 1D spectra extracted from the MUSE datacube using the segmentation map for each of the high confidence LAEs.
The shape of the emission line and the characteristic redshift compared to the systemic velocity of the LLS \citep{steidel2010,rak2011} strengthen our identification of these sources as LAEs.
As a final consistency check, we run an independent search of LAEs near the LLS using the {\sc LSDCat} tool \citep{herenz2017} based on a 3D match filtering and, consistently with the {\sc CubEx} analysis, we identify the same three high-confidence sources. 

\section{The environment of the $\lowercase{z}\approx 3.53$ LLS}\label{sec:thisdla}

The three LAEs detected at high-confidence are all within $\lesssim 200~\rm km~s^{-1}$ of the LLS with impact parameters ranging from $\approx 120$~kpc to $\approx 185$~kpc. The somewhat large offset in projection makes it unlikely that any of these galaxies are a direct counterpart of the absorber. However, the fact that these sources are found in a $\pm 1000~\rm km~s^{-1}$ window but all cluster at a much lower velocity offset is indicative of a physical association between the absorbing gas and the galaxies. In fact, offsets of this amplitude are readily explained by radiative transfer effects, whereby the Ly$\alpha$ emission line is typically redshifted by $\sim$100-300 km/s relative to systemic \citep{steidel2010,rak2011}, or peculiar velocities rather than the Hubble flow.  Thus, we conclude that our observations are likely revealing a galaxy-rich environment near the LLS.

To assess this statement more quantitatively, we evaluate the expected number of sources at these redshifts in a comparable search volume. The comoving volume defined by the MUSE FOV and the search window of $\pm 1000~\rm km~s^{-1}$ is $\approx 100~\rm Mpc^{3}$ at $z\approx 3.5$.  From the $z\approx 3-4$ field luminosity function \citep{grove2009,cassata2011,drake2017,herenz2019}, we would expect to find $\approx 0.3-1$ sources (depending on the assumed luminosity function, with a mean of $\approx 0.66$) for a luminosity of $\ge 4.5\times 10^{41}~\rm erg~s^{-1}$ where we are $\approx 50\%$ complete for extended sources.  Therefore the detection of three LAEs at the redshift of the LLS which is $\approx 5$ times above the expected mean random number even without folding in incompleteness, confirms that the absorbing system lies within an overdensity of galaxies. To reinforce this argument, we perform an identical search within two $\pm 1000~\rm km~s^{-1}$ windows at redshifts above and below that of the LLS, where there are no absorption systems seen in the quasar spectrum. In both of these windows, we find no high-confidence LAEs consistent with the estimate above.

Having established that a galaxy-rich environment exists near the $z\approx 3.53$ LLS, we now turn to possible scenarios for the nature of this absorption system.
Recent studies of LLSs at $z\gtrsim 2$  have determined that they show a wide distribution of metallicity, covering four orders of magnitude with a peak at $\log(Z/Z_{\odot})\approx -2$ \citep{cooper2015,fumagalli2016,lehner2016}. However, only $\approx 10$ per cent are expected to have metallicities below $\log(Z/Z_{\odot}) \approx -3$. While some LLSs have been discovered with no detectable metals, e.g. LLS1134  with a metallicity of $\log(Z/Z_{\odot})\lesssim -4.2$ and LLS0956B  with $\log(Z/Z_{\odot}) \lesssim -3.8$ in \citet{fumagalli2011sci} and LLS1723 with $\log(Z/Z_{\odot})\lesssim -4.14$ \citep{robert2019}, the LLS in this field has one of the lowest confirmed metallicities to date, with $\log(Z/Z_{\odot}) = -3.41 \pm 0.26$ \citep{crighton2016}.

While the origin of these types of metal-poor structures is far from clear, various scenarios have been put forward. Combining the observed properties in absorption with those of the environment probed in emission, we can offer some additional insight into the nature of this LLS. In the case of LLS1249, its very low metallicity appears to rule out that the LLS is a galactic wind from one of the detected galaxies (or from undetected systems at closer impact parameter).  Stellar feedback (or even single supernovae events) within a host galaxy would likely enrich the LLS above the observed levels even when allowing for some mixing and dilution in metal-poor gas \citep[e.g.][]{wise2012,creasey2015,hafen2019}.

Based on the absorption properties alone, \citet{crighton2016} considered a Population II enrichment scenario where the LLS lies within a cold stream or low mass halo. The cold stream scenario was thought to be unlikely as the photoionisation models favoured larger cloud sizes than the few kpc expected for such structures. Similarly, the low-mass halo was also ruled out as the velocity width of the LLS is much larger than the virial velocity of a low mass halo ($10^{10}~\rm M_{\odot}$) at z$\sim$3.
Instead, \citet{crighton2016} favoured a scenario in which the LLS lies in the IGM and is thus not directly associated with galaxies. Under this scenario, the LLS would arise from a pristine region in the early Universe which collapsed to create Population III stars. These stars were the sole source of metal enrichment for the cloud and supernova events from these early stars produced feedback that stopped the accretion of further gas. As a result, after this early epoch the gas cloud produced few new stars and remained in the IGM with no interaction with galaxies to further enrich its gas. 

However, while our analysis of the \qname\ sightline has not detected any continuum sources in the environment of LLS1249, indicating that the LLS is not hosted by a highly (unobscured) star-forming galaxy, we have detected three LAEs. These sources are found at impact parameters of $\lesssim 185$~kpc with the closest at $\lesssim 120$~kpc, and all within $200~\rm km~s^{-1}$ of the systemic redshift of the LLS.
With Ly$\alpha$ luminosities ranging between $3.3\times10^{41}~\rm erg~s^{-1}$  and $3.5\times10^{42}~\rm erg~s^{-1}$, these LAEs are forming stars at a rate of $\approx 0.3-3.5~\rm M_{\odot}~yr^{-1}$, assuming the conversion factor in \citet{furlanetto2005} and not including dust extinction. Due to the scattering of Ly$\alpha$ photons, this is likely to be a lower limit on the true SFR. 
Based on the clustering analysis of LAEs \citep{gawiser2007, ouchi2010, bielby2016}, these sources are likely to live in halos of masses $\approx 10^{10.9 - 11.4}~ \rm M_{\odot}$, for which the corresponding virial radii at $z\approx 3.5$ is $\approx30 -45~\rm kpc$. Thus, while the LLS is not in the inner CGM of the LAEs at the time of observation, it still lies in the sphere of influence of these galaxies where relics of outflows travelling at $\approx 200-300~\rm km~s^{-1}$ can still raise the metal content on scales of $\approx 200-250~\rm kpc$ in $\approx 1~\rm Gyr$ (e.g. \citet{diaz2015}. Moreover, due to the observed clustering of galaxies near this LLS, it is likely that additional fainter galaxies reside in this region, potentially contributing to the enrichment \citep[e.g.][]{booth2012}. 

Based on these new observations, we can refine the picture for the origin of LLS1249 now the galaxy environment is considered. The discovery of LAEs in the MUSE data calls for a revision of the scenario proposed in \citet{crighton2016} which was based on absorption alone, in which we would expect the LLS to be found in a mostly underdense region and not associated with galaxies. Given the presence of multiple LAEs, a plausible scenario is that the gas originates in a structure connected to halos. Within numerical simulations, metal poor gas pockets in proximity of galaxies are often found in gas filaments connecting and feeding galaxies \citep[e.g.][]{faucher2011,fumagalli2011,vandevoort2012}. Within these filaments, multiple low-mass galaxies are often found \citep[e.g.][]{shen2012}, which would explain a rich environment like the one probed by our observations. 
A more detailed comparison with cosmological simulations that capture the  link between LLSs and LAEs is however required to model further these observations.  

Empirically,  due to its near-pristine enrichment, this gas may have been solely polluted by Population III stars as proposed originally by \citet{crighton2016}. As discussed above, however, the presence of star-forming galaxies in close proximity makes still plausible a Population II enrichment due to relics of outflows from AGNs or supernovae at higher redshift, which have been substantially diluted with cosmic time. It is also clear from the presence of such low metallicity gas within the environment of multiple LAEs that metal enrichment at these redshifts is still inhomogeneous on physical scales of $\approx 200~\rm kpc$, in line with the enrichment pattern often seen in different velocity components of individual LLSs \citep[e.g.][]{prochter2010}. Finally, as this gas structure lies within a galaxy overdensity, with time, this gas is likely to be accreted onto a nearby galaxy fuelling its star formation, for instance in the form of cold streams \citep[e.g.][]{keres2005,dekel2009}.

It is also interesting to note the analogy between LLS1249 and LLS0956B, another very-metal poor LLS studied with MUSE \citep{fumagalli2016}. In both cases, gas with metallicity $\log(Z/Z_{\odot}) \lesssim -3.5$
is seen in close proximity to multiple galaxies, in line with the prediction that relative metal-poor gas streams feed galaxy formation at these redshifts. In contrast, another LLS (LLS0956A) in the same sightline with $\log(Z/Z_{\odot}) = -3.35 \pm 0.05$ shows no galaxy associations down to the same sensitivity limit as our survey, favouring a more isolated IGM environment for very-metal poor LLSs.
Given the low number of sources studied to date, it is difficult to resolve which one of the two scenarios is more characteristic for the population of metal-poor LLSs.
A larger study of the environments of LLSs, which will be provided by the full MAGG survey, will therefore be crucial to obtain a statistical view of the nature of these types of systems, so as to build a more complete understanding of their origin.

\section{Summary and Survey goals}\label{sec:summary}

In this work, we have presented the MUSE Analysis of Gas around Galaxies (MAGG) survey together with results from the analysis of the first sightline. The MAGG survey is a 106-hour large programme on the MUSE instrument at the VLT that has been designed to investigate the connection between optically-thick gas and galaxies at $z \sim 3-4$. Our sample comprises 28 quasars at $z\gtrsim 3.2$, all with high-resolution and high S/N archival spectroscopy revealing at least one strong absorption line system along the line of sight. In total, these sightlines include \nlls\ strong absorption line systems at $z\gtrsim 3$.
In this paper, we have discussed in detail the methodology adopted for the reduction and analysis of both the MUSE and high-resolution spectroscopic data of the quasars, with particular emphasis on the post-processing of MUSE data and the techniques adopted for extracting continuum sources and line emitters from the data cubes.

The sightline studied in this first paper, \qname, contains a LLS at $z\approx 3.53$. This system was studied in detail by \citet{crighton2016} who determined, for a component with multiple detected ions, a metallicity of $\rm [Si/H] = -3.41 \pm 0.26$. This determination makes LLS1249 one of the most metal-poor LLSs yet known, and a possible candidate for Population III enrichment. To investigate the origin of this type of system and its relation to galaxies, we have used MUSE observations to search for galaxies in the environment of this absorption system. We have not found any continuum sources that could be associated with the LLS, but we have detected three high-confidence LAEs with projected separations less than $185~\rm kpc$ and velocities within $\lesssim 200~\rm km~s^{-1}$ of the LLS. 

The presence of these star-forming galaxies, within the environment of one of the most metal-poor structures known, indicates that metal enrichment is inhomogeneous at these redshifts, and provides new clues to the origin of extremely metal-poor gas clouds. Combining absorption spectroscopy and MUSE data, we have revised the proposed formation scenario in which LLS1249 is part of the IGM and was formed in an underdense region of the early Universe that has been solely polluted by Population III stars. Our results instead suggest the LLS1249 is part of a gas structure, possibly a filament, accreting onto galaxies. This gas has been enriched either by a Population III episode or by relics of high-redshift Population II outflows that have been diluted with cosmic time. As it resides in an overdensity, this gas is likely to be accreted with time onto a galaxy further fuelling star formation in this region, in analogy with what is seen in a similar LLS at this redshift.

While this analysis has once again highlighted the power of combining information in absorption with knowledge of the galaxy environment in emission, the small sample of sightlines presented to date in the literature prevents far-reaching conclusions on the nature of optically-thick gas clouds at $z\gtrsim 3$. Through the analysis of 27 further fields with $\gtrsim 50$ LLSs, the MAGG survey will enable us to conduct a large and unbiased search for galaxies clustered to strong absorption line systems. This search is expected to yield the detection of tens of star-forming galaxies close to these quasar sightlines, thus creating a novel dataset for statistical studies of the gas environment around galaxies and the connection between gas and star formation in the first $2~\rm Gyr$ of cosmic history. This will enable us to finally pin down the relative importance of different scenarios in the formation of LLSs spanning a wide range of column densities and metallicities.

\section*{Acknowledgements}
The authors thank N. Crighton, J. Hennawi, and M. Swinbank for their contributions to the design of this project. MF acknowledges support by the Science and Technology Facilities Council [grant number  ST/P000541/1]. This project has received funding from the European Research Council (ERC) under the European Union's Horizon 2020 research and innovation programme (grant agreement No 757535). MTM thanks the Australian Research Council for Discovery Project grant DP130100568 which supported this work. SC gratefully acknowledges support from Swiss National Science Foundation grant PP00P2\_163824. During this work, RJC was supported by a Royal Society University Research Fellowship.
This work is based on observations collected at the European Organisation for Astronomical Research in the Southern Hemisphere under ESO programmes ID 
197.A-0384, 
065.O-0299,
067.A-0022,
068.A-0461,
068.A-0492,
068.A-0600,
068.B-0115,
069.A-0613,
071.A-0067,
071.A-0114,
073.A-0071,
073.A-0653,
073.B-0787,
074.A-0306,
075.A-0464,
077.A-0166,
080.A-0482,
083.A-0042,
091.A-0833,
092.A-0011,
093.A-0575,
094.A-0280,
094.A-0131,
094.A-0585,
095.A-0200,
096.A-0937,
097.A-0089,
099.A-0159,
166.A-0106,
189.A-0424.
This work used the DiRAC Data Centric system at Durham University, operated by the Institute for Computational Cosmology on behalf of the STFC DiRAC HPC Facility (www.dirac.ac.uk). This equipment was funded by BIS National E-infrastructure capital grant ST/K00042X/1, STFC capital grants ST/H008519/1 and ST/K00087X/1, STFC DiRAC Operations grant ST/K003267/1 and Durham University. DiRAC is part of the National E-Infrastructure. This research made use of Astropy, a community-developed core Python package for Astronomy \citep{astropy}. Raw and reduced MUSE data are distributed via the ESO archive at \url{http://archive.eso.org/}.
This research has made use of the NASA/IPAC Extragalactic Database (NED) which is operated by the Jet Propulsion Laboratory, California Institute of Technology, under contract with the National Aeronautics and Space Administration.
This research has made use of data from the Sloan Digital Sky Survey IV (\url{www.sdss.org}), funded by the Alfred P. Sloan Foundation, the U.S. Department of Energy Office of Science, and the Participating Institutions. 
Some of the data presented herein were obtained at the W. M. Keck Observatory, which is operated as a scientific partnership among the California Institute of Technology, the University of California and the National Aeronautics and Space Administration. The Observatory was made possible by the generous financial support of the W. M. Keck Foundation. This research has made use of the Keck Observatory Archive (KOA), which is operated by the W. M. Keck Observatory and the NASA Exoplanet Science Institute (NExScI), under contract with the National Aeronautics and Space Administration. The authors wish to recognise and acknowledge the very significant cultural role and reverence that the summit of Maunakea has always had within the indigenous Hawaiian community.  We are most fortunate to have the opportunity to conduct observations from this mountain. 
Some of the data presented in this work were obtained from the Keck Observatory Database of Ionized Absorbers toward QSOs (KODIAQ), which was funded through NASA ADAP grant NNX10AE84G. The Keck Observatory Archive (KOA) is a collaboration between the NASA Exoplanet Science Institute (NExScI) and the W. M. Keck Observatory (WMKO). NExScI is sponsored by NASA's Exoplanet Exploration Program, and operated by the California Institute of Technology in coordination with the Jet Propulsion Laboratory (JPL). For access to other data products and codes used in this work, please contact the authors or visit \url{http://www.michelefumagalli.com/codes.html}.







\bsp 
\label{lastpage}

\end{document}